\documentclass[%
  a4paper, twocolumn, superscriptaddress, amsmath, amssymb, floatfix
]{revtex4-1}
\usepackage{amsthm}
\usepackage{graphicx}
\usepackage{epstopdf}
\usepackage[table]{xcolor}
\usepackage{braket}
\usepackage[colorlinks,bookmarksopen=false]{hyperref}
\usepackage[caption=false]{subfig}
\usepackage{bbm}
\usepackage{makecell}
\usepackage{lipsum, blindtext}
\usepackage{xr}
\usepackage{MnSymbol}
\usepackage{siunitx}
\usepackage[normalem]{ulem}
\usepackage{cancel}

\hypersetup{%
  pdfstartview={FitW},
  linkcolor=blue,
  citecolor=blue,
  filecolor=black,
  menucolor=black,
  urlcolor =blue
}

\newcommand{\im}{\ensuremath{\textup{i}}}

\newcommand{\rp}{\ensuremath{\mathfrak{Re}}}
\newcommand{\ip}{\ensuremath{\mathfrak{Im}}}

\newcommand{\op}[1]{\ensuremath{\mathsf{#1}}}
\newcommand{\lop}[1]{\ensuremath{\mathcal{#1}}}
\newcommand{\vecop}[1]{\ensuremath{\boldsymbol{\mathsf{\hat{#1}}}}}
\newcommand{\uop}{\op{\mathbbm{1}}}

\newcommand{\error}{\varepsilon}
\newcommand{\pulse}{\mathcal{E}}
\newcommand{\dmap}{\mathcal{D}}

\newcommand{\hil}{\mathcal{H}}

\newtheorem*{mydef*}{Definition}
\newtheorem*{myprop*}{Proposition}

\newtheorem*{mythm*}{Theorem}

\newcommand{\myvec}[1]{\boldsymbol{#1}}

\newcommand{\dd}{\ensuremath{\mathrm{d}}}

\newcommand{\tr}{\ensuremath{\mathrm{tr}}}

\newcommand{\lol}[1]{%
  \left\llangle#1\right\rrangle
}

\newcommand{\rmtrgt}{\mathrm{trgt}}
\newcommand{\rmgrad}{\mathrm{grad}}
\newcommand{\rmeff}{\mathrm{eff}}
\newcommand{\rmopt}{\mathrm{opt}}
\newcommand{\rmref}{\mathrm{ref}}
\newcommand{\rmBS}{\mathrm{BS}}
\newcommand{\rmAB}{\mathrm{AB}}
\newcommand{\rmsb}{\mathrm{sb}}
\newcommand{\rmin}{\mathrm{in}}
\newcommand{\rma}{\mathrm{a}}
\newcommand{\rmb}{\mathrm{b}}
\newcommand{\rmc}{\mathrm{c}}
\newcommand{\rmL}{\mathrm{L}}

\newcommand{\rmx}{\mathrm{x}}
\newcommand{\rmz}{\mathrm{z}}

\newcommand{\som}{\tilde{\omega}}

\newcommand{\dket}[1]{\ket{\overline{#1}}}
\newcommand{\dbra}[1]{\bra{\overline{#1}}}

\newcommand{\na}{n_{\rma}}
\newcommand{\nb}{n_{\rmb}}
\newcommand{\nc}{n_{\rmc}}

\begin{document}

\title{%
  Engineering Strong Beamsplitter Interaction between Bosonic Modes via Quantum
  Optimal Control Theory
}

\author{Daniel Basilewitsch}
\altaffiliation[present address: ]{%
  Institute for Theoretical Physics, University of Innsbruck, A-6020
  Innsbruck, Austria
}
\affiliation{%
  Dahlem Center for Complex Quantum Systems and Fachbereich Physik, Freie
  Universit\"{a}t Berlin, D-14195 Berlin, Germany
}
\affiliation{%
  Theoretische Physik, Universit\"{a}t Kassel, D-34132 Kassel, Germany
}
\affiliation{%
  Yale Quantum Institute, Yale University, New Haven, Connecticut 06511, USA
}

\author{Yaxing Zhang}
\author{S. M. Girvin}
\affiliation{%
  Yale Quantum Institute, Yale University, New Haven, Connecticut 06511, USA
}
\affiliation{%
  Department of Physics, Yale University, New Haven, Connecticut 06511, USA
}

\author{Christiane P. Koch}
\email{christiane.koch@fu-berlin.de}
\affiliation{%
  Dahlem Center for Complex Quantum Systems and Fachbereich Physik, Freie
  Universit\"{a}t Berlin, D-14195 Berlin, Germany
}
\affiliation{%
  Theoretische Physik, Universit\"{a}t Kassel, D-34132 Kassel, Germany
}

\date{\today}

\begin{abstract}
  In continuous-variable quantum computing with qubits encoded in the
  infinite-dimensional Hilbert space of bosonic modes, it is a difficult task to
  realize strong and on-demand interactions between the qubits. One option is to
  engineer a beamsplitter interaction for photons in two superconducting
  cavities by driving an intermediate superconducting circuit with two
  continuous-wave drives, as demonstrated in a recent experiment
  [\href{https://link.aps.org/doi/10.1103/PhysRevX.8.021073}{Gao \emph{et al.}, Phys. Rev. X \textbf{8}, 021073 (2018)}].
  Here, we show how quantum optimal control theory (OCT) can be used in
  a systematic way to improve the beamsplitter interaction between the two
  cavities. We find that replacing the two-tone protocol by a three-tone
  protocol accelerates the effective beamsplitter rate between the two cavities.
  The third tone's amplitude and frequency are determined by gradient-free
  optimization and make use of cavity-transmon sideband couplings. We show how
  to further improve the three-tone protocol via gradient-based optimization
  while keeping the optimized drives experimentally feasible. Our work
  exemplifies how to use OCT to systematically improve practical protocols in
  quantum information applications.
\end{abstract}

\maketitle

\section{Introduction}
Quantum technologies~\cite{Acin2018} such as quantum computing~\cite{Nielsen},
quantum simulation~\cite{RMP.86.153} or quantum sensing~\cite{RMP.89.035002}
promise to outperform their classical analogues by exploiting quantum properties
like coherence and entanglement. A high degree of control over the underlying
quantum systems is required for their practical realization, since operating a
quantum device implies the capability to steer the system's dynamics in the
desired way. Electromagnetic fields, which interact with the quantum system and
which can be shaped in time, are typical control knobs. Unfortunately, deriving
suitable field shapes quickly becomes non-trivial for increasing complexity of
either the quantum system or the control problem~\cite{DAlessandro}. Optimal
control theory (OCT) has developed around this non-trivial
task~\cite{EPJD.69.279}, providing tools to calculate the field shapes needed to
obtain a desired dynamics, e.g.\ with smallest error or in shortest time. While
OCT for quantum control was first applied in the context of NMR~\cite{Mao1986,
Murdoch1987} and molecular dynamics~\cite{Tannor1985, PRA.37.4950, Kosloff1989},
OCT has more recently been attracting attention in the field of quantum
technologies. This entailed significant method development, concerning both
optimization targets~\cite{PRA.84.042315, WattsPRA2015} and optimization
algorithms~\cite{CanevaPRA2011, Goerz2015, PRAppl.15.034080} to ease
implementation of constraints ensuring experimental feasibility. To this end,
tailored optimization algorithms~\cite{Skinner2010, CanevaPRA2011,
SorensenPRA2018, MachnesPRL2018, Gunther2021} have been developed, which only
explore a restricted function space for solutions but yield smooth field shapes.
Alternatively, gradient-based optimization techniques~\cite{JMagRes.172.296,
SciPostPhys.7.6.080} can be used, which explore an unrestricted function space
but might require additional constraints~\cite{PRA.77.063412, PRA.88.053409} to
keep the field shapes smooth and feasible. A hybrid optimization approach which
combines gradient-free and gradient-based techniques is another option combining
advantages from both methods~\cite{Goerz2015}. It pre-selects promising field
shapes via gradient-free optimization --- exploring only a small function space
--- and fine-tunes these fields afterwards via gradient-based methods. By now,
OCT has become a versatile and reliable tool that delivers solutions for the
various control problems across sub-disciplines of quantum
physics~\cite{EPJD.69.279, Koch2016}.

The utility of OCT in the field of quantum technologies is confirmed by
successful application in various experiments, for instance to improve the
performance of protocols for quantum computing~\cite{Dolde2014, Waldherr2014,
Heeres2017, PRL.125.170502, Werninghaus2021}, quantum
simulation~\cite{Omran2019} and quantum sensing~\cite{PRL.115.190801,
PRX.8.021059, PRX.10.021058, Titum2021}. While these advances are impressive,
use of optimized pulses in experiment typically involves significant seesaw of
improving experimental calibration and theoretical fine-tuning of the pulses.
Lack of intelligibility of brute-force optimized pulses, as obtained from e.g.\
gradient-based techniques, often further hampers this process. A viable route
from OCT to laboratory application is therefore still missing. Here, we argue
that hybrid optimization~\cite{Goerz2015} provides a systematic way to design
intelligible and experimentally feasible pulses, using a practical problem,
relevant for continuous-variable quantum computing as example. In particular,
pre-optimization with a reduced number of control parameters facilitates the
derivation of intelligible control solutions. These can then be brought to
maximal performance in the second stage of optimization.

Continuous-variable quantum computing~\cite{PRL.80.4084, RevModPhys.77.513} is
a promising approach for building a quantum computer~\cite{Nielsen}, harnessing
the infinite-dimensional Hilbert spaces of bosonic modes to encode and process
quantum information~\cite{PRA.64.012310}. This may provide an advantage over
quantum information platforms with finite-dimensional Hilbert spaces when it
comes to quantum error correction~\cite{Joshi2021}. While noise-protection is
a challenging task for traditional qubit platforms such as superconducting
circuits~\cite{Gyenis2021}, substantial progress has been made in recent years
in protecting bosonic modes~\cite{Terhal2020}. This encompasses the proposal of
new error-correction codes~\cite{Mirrahimi2014, PRX.6.031006, Albert2019} as
well as recent experimental demonstrations~\cite{Ofek2016, Grimm2020, Hu2019,
Gertler2021} of such codes, making bosonic modes an attractive platform to
achieve universal, error-corrected quantum computation.

The capability to entangle qubits on-demand is one important prerequisite ---
among others~\cite{FortschrPhys.48.771} --- for any successful quantum computing
platform. It requires a controlled interaction between the qubits. While the
implementation of entangling gates is nowadays carried out rather routinely
between e.g.\ superconducting circuits~\cite{PRL.109.240505, PRL.109.060501} or
trapped ions~\cite{PRL.117.060505, PRL.117.060504}, it is still a non-trivial
task for qubits in continuous-variable settings~\cite{PRL.92.123601,
Masada2015}. In recent years, hybrid approaches to continuous-variable quantum
computing which combine elements like superconducting
cavities~\cite{Reagor2013}, to host the bosonic modes, with elements from
circuit quantum electrodynamics, have been investigated~\cite{Joshi2021,
Ma2021}. Interestingly, these hybrid approaches reverse the roles of cavities
and circuits compared to the more traditional protocols employing
superconducting circuits as qubits~\cite{Gambetta2017, Kjaergaard2020}. In
contrast, in the hybrid approach, the cavities are controlled via
superconducting circuits~\cite{Ma2021}, e.g.\ by using optimized pulses on the
control circuits~\cite{Heeres2017}. This allows for new ways to let the
cavities, i.e., the qubits, interact and thus realize entangling gates. For the
latter, however, it matters how the qubits are encoded within each cavity. In
other words, the entangling protocol depends on how the qubit's two logical
basis states are encoded within the infinite-dimensional Hilbert space of the
bosonic modes. To take advantage of the excellent error-correction capabilities
of bosonic modes, the two logical basis states are typically specifically
selected to be less susceptible to decoherence~\cite{Mirrahimi2014,
PRX.6.031006}. For two superconducting cavities interacting via an intermediate
transmon qubit, feasibility of entangling operations for specific encodings has
been demonstrated recently~\cite{Rosenblum2018, Chou2018}. Using the same setup,
a codeword-agnostic solution, i.e., an entangling gate that works for any
encoding, has been demonstrated shortly after~\cite{Gao2019}. This
codeword-agnostic gate depends on an engineered beamsplitter interaction between
the two cavities that can be activated on-demand by driving the intermediate
transmon qubit by two continuous-wave drives~\cite{PRX.8.021073}.

Here, we use the setup and protocol of Ref.~\cite{PRX.8.021073} and show how to
enhance the engineered beamsplitter interaction between two cavities using the
hybrid optimization approach introduced in Ref.~\cite{Goerz2015}. We demonstrate
that extending the protocol of Ref.~\cite{PRX.8.021073} --- in the following
called two-tone protocol --- by a third continuous-wave drive --- in the
following called three-tone protocol --- leads to an increase of the effective
beamsplitter interaction strength. We explain how the choice of the third
drive's amplitude and frequency --- determined by gradient-free optimization ---
can be understood. To this end, we show analytically that the enhancement of the
beamsplitter interaction comes from the third drive's frequency being chosen by
the algorithm such as to create near-resonant sideband couplings between both
cavities and the transmon. The two-tone and three-tone protocols can be further
improved using gradient-based optimization --- in the following called
fine-tuned two- and three-tone protocol ---, while keeping the optimized drives
feasible. We furthermore discuss the impact of decoherence on all protocols and
how the errors change as the coherence times improve. Our work exemplifies how
to use OCT to obtain intelligible and feasible solutions to practical problems
in quantum technologies.

The paper is organized as follows. In Sec.~\ref{subsec:model} we introduce the
physical model and the control strategy employed in Ref.~\cite{PRX.8.021073}. In
the subsequent Secs.~\ref{subsec:trgts} and~\ref{subsec:oct} we introduce the
control problem that we want to solve and give a brief overview of the technical
aspects of OCT. Section~\ref{sec:strong} presents our main results. While in
Secs.~\ref{subsec:2vs3} and~\ref{subsec:grad} we present our control solution
and how it can be found using numerical methods, in Sec.~\ref{subsec:ana} we
explain the physical mechanism behind the solution using analytical tools.
Section~\ref{sec:diss} compares the performance of our solution with that of
Ref.~\cite{PRX.8.021073} in the presence of decoherence.
Section~\ref{sec:conclusions} concludes.

\section{Model and Methods}

\subsection{Model}
\label{subsec:model}
\begin{figure}[tb!]
  \centering
  \includegraphics{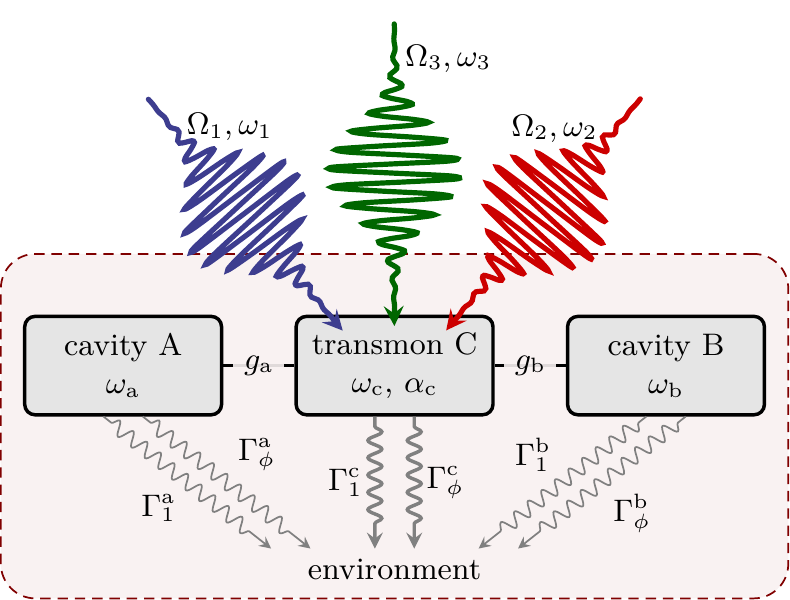}
  \caption{%
    Tripartite system consisting of two superconducting cavities, labeled A and
    B, coupled via an intermediate (driven) transmon, labeled C. The system
    interacts with its environment, modeled by individual decay rates for
    each subsystem.
  }
  \label{fig:model}
\end{figure}

We consider a tripartite system, sketched in Fig.~\ref{fig:model}, consisting of
two superconducting cavities~\cite{Reagor2013}, labeled A and B, which both
couple to an intermediate transmon qubit~\cite{PRA.76.042319}, labeled C. The
cavity modes are modeled by harmonic oscillators while the transmon is given
by an anharmonic oscillator. In the lab frame, the Hamiltonian
reads~\cite{PRA.99.012314}
\begin{align} \label{eq:ham}
  \op{H}(t)
  &=
  \omega_{\rma} \op{a}^{\dagger} \op{a}
  +
  \omega_{\rmb} \op{b}^{\dagger} \op{b}
  +
  \omega_{\rmc} \op{c}^{\dagger} \op{c}
  -
  \frac{\alpha_{\rmc}}{2} \op{c}^{\dagger} \op{c}^{\dagger} \op{c} \op{c}
  \notag \\
  &\quad
  +
  g_{\rma} \left(\op{a} \op{c}^{\dagger} + \op{a}^{\dagger} \op{c}\right)
  +
  g_{\rmb} \left(\op{b} \op{c}^{\dagger} + \op{b}^{\dagger} \op{c}\right)
  \notag \\
  &\quad
  +
  \sum_{k} \left(%
    \Omega_{k}(t) e^{- \im \omega_{k} t} \op{c}^{\dagger}
    +
    \Omega_{k}^{*}(t) e^{\im \omega_{k} t} \op{c}
  \right),
  \vphantom{\frac{1}{2}}
\end{align}
where $\op{a}, \op{b}$ and $\op{c}$ are the annihilation operators for the modes
of cavities A and B and transmon C, respectively. $\omega_{\rma}$ and
$\omega_{\rmb}$ are the frequencies of the two cavity modes and $\omega_{\rmc}$
corresponds to the frequency difference between the ground and first excited
state of the transmon C. $\alpha_{\rmc} \ll \omega_{\rmc}$ describes the
transmon's anharmonicity for higher level splittings. $g_{\rma}$ and $g_{\rmb}$
are the static couplings between the cavity A/B and transmon C, respectively.
Note that doubly exciting (de-exciting) terms like $\op{a}^{\dagger}
\op{c}^{\dagger}$ ($\op{a} \op{c}$) and $\op{b}^{\dagger} \op{c}^{\dagger}$
($\op{b} \op{c}$) have been neglected. The last row in Eq.~\eqref{eq:ham}
describes the interaction of a set of control fields with transmon C, with
$\Omega_{k}(t)$ and $\omega_{k}$ the time-dependent amplitude and frequency of
field $k$. In addition, we account for the interaction of the tripartite system
with its environment and model the environment's influence via
a Gorini-Kossakowski-Sudarshan-Lindblad master equation~\cite{Breuer},
\begin{align} \label{eq:LvN}
  \frac{\dd}{\dd t} \op{\rho}(t)
  &=
  - \im \left[\op{H}(t), \op{\rho}(t)\right]
  \notag \\
  &\quad
  +
  \sum_{p,\rmx}
  \Gamma_{p}^{\rmx} \Big(%
    \op{L}_{p}^{\rmx}
    \op{\rho}(t)
    \op{L}_{p}^{\rmx} \dagger
    - \frac{1}{2} \big\{%
      \op{L}_{p}^{\rmx} \dagger
      \op{L}_{p}^{\rmx}, \op{\rho}(t)
    \big\}
  \Big)
  \notag \\
  &=
  \lop{L}(t) \left[\op{\rho}(t)\right].
\end{align}
The Lindblad operators $\op{L}_{p}^{\rmx}$ and their corresponding decay rates
$\Gamma_{p}^{\rmx}$ are chosen such as to describe relaxation and pure dephasing
processes on each individual subsystem $\rmx = \rma, \rmb, \rmc$, i.e.,
\begin{align} \label{eq:lb_ops}
  &\op{L}_{1}^{\rma}
  =
  \op{a},
  &&\op{L}_{2}^{\rma}
  =
  \op{a}^{\dagger} \op{a},
  &&\Gamma_{1}^{\rma}
  =
  \frac{1}{T_{1}^{\rma}},
  &&\Gamma_{2}^{\rma}
  =
  \frac{2}{T_{\phi}^{\rma}},
  \notag \\
  &\op{L}_{1}^{\rmb}
  =
  \op{b},
  &&\op{L}_{2}^{\rmb}
  =
  \op{b}^{\dagger} \op{b},
  &&\Gamma_{1}^{\rmb}
  =
  \frac{1}{T_{1}^{\rmb}},
  &&\Gamma_{2}^{\rmb}
  =
  \frac{2}{T_{\phi}^{\rmb}},
  \notag \\
  &\op{L}_{1}^{\rmc}
  =
  \op{c},
  &&\op{L}_{2}^{\rmc}
  =
  \op{c}^{\dagger} \op{c},
  &&\Gamma_{1}^{\rmc}
  =
  \frac{1}{T_{1}^{\rmc}},
  &&\Gamma_{2}^{\rmc}
  =
  \frac{2}{T_{\phi}^{\rmc}},
\end{align}
where $T_{1}^{\rmx}$ and $T_{\phi}^{\rmx}$ are the individual $T_{1}$ relaxation
and $T_{\phi}$ pure dephasing times. Note that for numerical efficiency, we work
in a rotating frame, where Hamiltonian~\eqref{eq:ham} becomes
\begin{align} \label{eq:ham'}
  \op{H}'(t)
  &=
  -
  \frac{\alpha_{\rmc}}{2} \op{c}^{\dagger} \op{c}^{\dagger} \op{c} \op{c}
  \notag \\
  &\quad
  +
  g_{\rma} \left(%
    e^{- \im \delta_{\rma} t}
    \op{a} \op{c}^{\dagger}
    +
    e^{  \im \delta_{\rma} t}
    \op{a}^{\dagger} \op{c}
  \right)
  \notag \\
  &\quad
  +
  g_{\rmb} \left(%
    e^{- \im \delta_{\rmb} t}
    \op{b} \op{c}^{\dagger}
    +
    e^{  \im \delta_{\rmb} t}
    \op{b}^{\dagger} \op{c}
  \right)
  \notag \\
  &\quad
  +
  \sum_{k} \left(%
    \Omega_{k}(t) e^{- \im \delta_{k} t} \op{c}^{\dagger}
    +
    \Omega_{k}^{*}(t) e^{\im \delta_{k} t} \op{c}
  \right),
\end{align}
with $\delta_{\rma (\rmb)} = \omega_{\rma (\rmb)} - \omega_{\rmc}$ and
$\delta_{k} = \omega_{k} - \omega_{\rmc}$.

In the following, we want to engineer a beamsplitter interaction between the two
cavities by driving the intermediate transmon appropriately~\cite{PRX.8.021073},
i.e., we want to engineer a Hamiltonian of the form
\begin{align} \label{eq:HabBS}
  \op{H}_{\rmBS}^{\rmAB}(t)
  =
  g_{\rmBS}(t) \op{a} \op{b}^{\dagger}
  +
  g_{\rmBS}^{*}(t) \op{a}^{\dagger} \op{b}
\end{align}
on the reduced system of the two cavities. Here, $g_{\rmBS}(t)$ corresponds to
the effective interaction strength, i.e., the beamsplitter rate between cavities
A and B. One way to realize the interaction is to drive the transmon with two
control fields, $\Omega_{1}(t)$ and $\Omega_{2}(t)$, with constant frequencies,
$\omega_{1}$ and $\omega_{2}$~\cite{PRX.8.021073}. The latter need to fulfill
the resonance condition
\begin{align} \label{eq:reson}
  \som_{\rmb} - \som_{\rma}
  =
  \omega_{2} - \omega_{1},
\end{align}
where $\som_{\rma}$ and $\som_{\rmb}$ are the Stark-shifted versions of cavity
frequencies $\omega_{\rma}$ and $\omega_{\rmb}$, respectively. The individual
Stark shifts are induced by driving the transmon together with its small but
finite coupling to each cavity. Note that the Stark shifts are determined
primarily by the amplitudes $\Omega_{1}(t)$ and $\Omega_{2}(t)$ of the two
control fields and only weakly by their frequencies~\cite{PRA.99.012314}. To
ensure Eq.~\eqref{eq:reson}, the amplitudes should be kept constant --- except
for a small ramping time $\tau$ at the beginning and end of the protocol, in
order to switch the fields on and off smoothly. This can for instance be
achieved by choosing
\begin{align}
  \Omega_{k}(t)
  =
  \Omega_{k} S(t)
\end{align}
with the shape function
\begin{align} \label{eq:S}
  S(t)
  =
  \begin{cases}
    \sin^{2} \left(\pi \frac{t}{2 \tau}\right),
    &\qquad t \in [0, \tau),
    \\
    1,
    &\qquad t \in [\tau, T-\tau],
    \\
    \sin^{2} \left(\pi \frac{T-t}{2 \tau}\right),
    &\qquad t \in (T-\tau, T],
    \\
    0,
    &\qquad \mathrm{else},
  \end{cases}
\end{align}
where $T$ is the protocol's total duration.

In the following, we examine whether it is possible to increase the beamsplitter
rate, $|g_{\rmBS}(t)|$, by using an additional control field, respectively more
frequencies $\{\omega_{k}\}$, and by exploiting fully time-dependent amplitudes
$\{\Omega_{k}(t)\}$. To answer this question, i.e., to tackle this non-trivial
control task, we use quantum optimal control theory (OCT) to optimize
$\{\omega_{k}\}$ and $\{\Omega_{k}(t)\}$. Briefly, quantum control assumes
that a system can be steered by a set of control fields to a desired target.
Let $\{\pulse_{k}(t)\}$ be the set of fields for illustration purposes. OCT
provides the tools to derive tailored, i.e., optimized, control fields
$\{\pulse_{k}^{\rmopt}(t)\}$ realizing the corresponding dynamics, e.g.\
yielding the smallest error or shortest time~\cite{EPJD.69.279}.  We will
specify the physical aspects of the control problem in Sec.~\ref{subsec:trgts}
and the technical details on how to find optimized versions of $\{\omega_{k}\}$
and $\{\Omega_{k}(t)\}$ in Sec.~\ref{subsec:oct}.

\subsection{Target operations and encodings}
\label{subsec:trgts}
To tackle the control task of realizing a beamsplitter interaction with the
extended set of control fields, we must be able to quantify how well the
dynamics of the reduced system of the two cavities matches the desired one
generated by the ``target'' Hamiltonian~\eqref{eq:HabBS}. This is technically
challenging, since our ``figure of interest'' is not some accessible and
quantifiable feature of the dynamics but rather its generator. An ideal measure
would allow for a direct comparison of the target Hamiltonian with the
\emph{effective} Hamiltonian for the reduced system of the two cavities, given
the current choice of frequencies $\{\omega_{k}\}$ and amplitudes
$\{\Omega_{k}(t)\}$. Unfortunately, it is not possible to derive such an
expression for an effective Hamiltonian in case of an arbitrary (yet
unknown) choice of $\{\omega_{k}\}$ and $\{\Omega_{k}(t)\}$. However, we can
compare the dynamics, which the target and actual Hamiltonian give rise to, by
means of comparing various time-evolved states and quantifying their distance
with respect to some desired outcome. In detail, we compute the dynamical map
$\dmap_{T,0}^{\rmtrgt}$ that the desired Hamiltonian~\eqref{eq:HabBS} gives rise
to for any initial state $\op{\rho}_{\rmin}$ and quantify the distance between
the desired outcome $\op{\rho}_{\rmtrgt}(T) = \dmap_{T,0}^{\rmtrgt}
[\op{\rho}_{\rmin}]$ and the actual time-evolved state $\op{\rho}(T)
= \dmap_{T,0} [\op{\rho}_{\rmin}]$. The dynamical map $\dmap_{T,0}$ of the
actual time-evolution depends on $\{\omega_{k}\}$ and $\{\Omega_{k}(t)\}$. In
practice, it is not necessary to evaluate the distance for any
$\op{\rho}_{\rmin}$ but only for a set of basis states spanning the subspace
within which we require accurate execution of the protocol. This subspace will
be the logical two-qubit subspace in the following. To specify the latter ---
and thus the set of states for which to evaluate $\op{\rho}(T) = \dmap_{T,0}
[\op{\rho}_{\rmin}]$ --- we first introduce $\{\dket{\na, \nb, \nc}\}$, the
eigenstate basis of the field-free Hamiltonian~\eqref{eq:ham}. The nomenclature
of the eigenstate $\dket{\na, \nb, \nc}$ is chosen identical to that of the Fock
state $\ket{\na, \nb, \nc}$ with which it has the largest overlap. Given the
eigenstate basis, the target dynamical map $\dmap_{T,0}^{\rmtrgt}$ yields the
time-evolution
\begin{align} \label{eq:dmap_trgt}
  \dmap_{T,0}^{\rmtrgt} \left[%
    \dket{\na, \nb, 0} \dbra{\na, \nb, 0}
  \right]
  =
  \dket{\nb, \na, 0} \dbra{\nb, \na, 0}
\end{align}
for all $\na, \nb=0,1,2,\dots$. In other words, $\dmap_{T,0}^{\rmtrgt}$ swaps
(up to some phase) the states of the two cavities, leaving the transmon
invariant.

We seek a $\dmap_{T,0}$ that yields the same outcome as in
Eq.~\eqref{eq:dmap_trgt} if the frequencies $\{\omega_{k}\}$ and drive
amplitudes $\{\Omega_{k}(t)\}$ are chosen appropriately. Let $\op{\rho}_{\na,
\nb, \nc} = \dket{\na, \nb, \nc} \dbra{\na, \nb, \nc}$. A measure that becomes
zero if and only if $\dmap_{T,0}$ reproduces the desired outcome of
Eq.~\eqref{eq:dmap_trgt} and is strictly larger otherwise is given by
\begin{align} \label{eq:JTM}
  \error
  =
  1 - \frac{1}{M^{2}}
  \sum_{\na,\nb=0}^{M-1} \lol{%
    \op{\rho}_{\nb,\na,0}
    \Big|
    \dmap_{T,0} \left[
      \op{\rho}_{\na,\nb,0}
    \right]
  }
\end{align}
with $\lol{\op{A} | \op{B}} = \tr\{\op{A}^{\dagger} \op{B}\}$ and $M$ the
maximal photon number in the cavities up to which the correct behavior of
Eq.~\eqref{eq:dmap_trgt} is being checked. Note that the perfect beamsplitter
interaction of Eq.~\eqref{eq:HabBS} always gives rise to a perfect swap of the
cavity states, i.e., $\error=0$ holds in case of $\dmap_{T,0}
= \dmap_{T,0}^{\rmtrgt}$ for arbitrarily large $M$, i.e., arbitrary large photon
numbers in the cavities. Furthermore note that we assume the transmon to be
initially in its ground state and --- since a perfect beamsplitter operation
would leave the transmon state unchanged --- require $\dmap_{T,0}$ to return the
transmon to its ground state at time $T$.

It is the fact that a perfect beamsplitter interaction always gives rise to
a perfect swap of the cavity states that makes it so appealing for continuous
variable quantum computing. Any protocol or gate would then work independent of
the qubits' encoding, i.e., independent of the states $\ket{0}_{\rmL}$ and
$\ket{1}_{\rmL}$ chosen to represent the two logical qubit levels. Instead of
evaluating Eq.~\eqref{eq:JTM} for large $M$, which requires the propagation of
$M^{2}$ initial states, we consider two different encodings and check whether
the desired dynamics can be observed in the corresponding logical two-qubit
subspace. To this end, we consider an encoding of the logical
qubit states in the cavity's two lowest Fock states, i.e., $\ket{0/1}_{\rmL}
= \ket{0/1}$. The logical two-qubit basis within the tripartite system of the
two cavities and the transmon is then given by
\begin{align} \label{eq:fockL}
  \ket{0,0,0}_{\rmL}
  &\equiv
  \dket{0,0,0},
  \notag \\
  \ket{0,1,0}_{\rmL}
  &\equiv
  \dket{0,1,0},
  \notag \\
  \ket{1,0,0}_{\rmL}
  &\equiv
  \dket{1,0,0},
  \notag \\
  \ket{1,1,0}_{\rmL}
  &\equiv
  \dket{1,1,0}.
\end{align}
In contrast to this rather simple encoding, we also consider a binomial
encoding~\cite{PRX.6.031006} in which case the logical qubit states are given by
\begin{align}
  \ket{0}_{\rmL}
  =
  \frac{\ket{0} + \ket{4}}{\sqrt{2}},
  \qquad
  \ket{1}_{\rmL}
  =
  \ket{2}
\end{align}
within each cavity. Thus, the logical two-qubit basis within the tripartite
system is given by
\begin{align} \label{eq:binomL}
  \ket{0,0,0}_{\rmL}
  &\equiv
  \frac{\dket{0,0,0} + \dket{0,4,0} + \dket{4,0,0} + \dket{4,4,0}}{2},
  \notag \\
  \ket{0,1,0}_{\rmL}
  &\equiv
  \frac{\dket{0,2,0} + \dket{4,2,0}}{\sqrt{2}},
  \notag \\
  \ket{1,0,0}_{\rmL}
  &\equiv
  \frac{\dket{2,0,0} + \dket{2,4,0}}{\sqrt{2}},
  \notag \\
  \ket{1,1,0}_{\rmL}
  &\equiv
  \dket{2,2,0}.
  \vphantom{\frac{1}{\sqrt{2}}}
\end{align}
To compute the error of the protocol, we evaluate
\begin{align} \label{eq:JT}
  \error
  =
  1 - \frac{1}{4}
  \sum_{\na,\nb=0}^{1} \lol{%
    \op{\rho}_{\nb,\na,0}^{\rmL}
    \Big|
    \dmap_{T,0} \left[
      \op{\rho}_{\na,\nb,0}^{\rmL}
    \right]
  },
\end{align}
with $\op{\rho}_{\na, \nb, \nc}^{\rmL} = \ket{\na, \nb, \nc}_{\rmL} \bra{\na,
\nb, \nc}_{\rmL}$ for the two encodings given in Eqs.~\eqref{eq:fockL}
and~\eqref{eq:binomL}.

\subsection{Quantum Optimal Control Theory}
\label{subsec:oct}
We now turn towards  OCT, where a control problem is typically converted into
the minimization of a cost function. The latter is given by the optimization
functional
\begin{equation} \label{eq:J}
  \begin{aligned}
    J\left[\left\{\pulse_{k}\right\}, \left\{\op{\rho}_{l}\right\}\right]
    &=
    \error \left[\left\{\op{\rho}_{l}(T)\right\}\right]
    \\
    &\quad+
    \int_{0}^{T} \dd t\, J_{t}\left[\left\{\pulse_{k}(t)\right\},
    \left\{\op{\rho}_{l}(t)\right\}, t\right],
  \end{aligned}
\end{equation}
consisting of the final-time functional $\error [\{\op{\rho}_{l}(T)\}]$, which
quantifies how well the dynamics reaches a desired target at final time $T$, and
an intermediate-time functional $J_{t}[\{\pulse_{k}(t)\}, \{\op{\rho}_{l}(t)\},
t]$, which captures additional time-dependent costs and constraints.
$\left\{\op{\rho}_{l}(t)\right\}$ is a set of time-evolved states where the
index $l$ distinguishes different initial states. The choice of $J$ --- with
$\error$ its most important part --- captures the goal of the control problem.
Searching for the control fields that minimize $J$, i.e., solve the control
problem, yields optimized fields $\{\pulse_{k}^{\rmopt}(t)\}$ that implement the
desired dynamics best.

In the following, we use OCT in order to find optimized control fields, i.e.,
time-dependent drive amplitudes $\{\Omega_{k}^{\rmopt}(t)\}$, such that they
minimize the error $\error$, Eq.~\eqref{eq:JT}. We achieve this in two steps.
In a first step, we consider time-independent amplitudes (up to the ramps) as in
the original protocol~\cite{PRX.8.021073} but we add a third control field with
amplitude $\Omega_{3}$ and frequency $\omega_{3}$ to generate the desired
dynamics, Eq.~\eqref{eq:dmap_trgt}, in shorter time $T$. In this case, $\error$
becomes a function of $\Omega_{3}$ and $\omega_{3}$ as well as the final time
$T$. We use the gradient-free Nelder-Mead optimization method~\cite{CJ.7.308} to
search for an optimized set of these three parameters that minimizes $\error$.
In a second step, we then fix the frequencies $\omega_{1}, \omega_{2},
\omega_{3}$ of the three control fields as well as the final time $T$ and allow
their amplitudes, $\Omega_{1}(t), \Omega_{2}(t)$ and $\Omega_{3}(t)$, to be
fully time-dependent to minimize $\error$ even further. We use Krotov's method
for this purpose and briefly summarize its main equations in the following.

Krotov's method~\cite{AutomRemContr.60.1427} is an iterative, gradient-based
optimization technique with guaranteed monotonic
convergence~\cite{JCP.136.104103}. In order to obtain an update equation for
each field $\Omega_{k}(t)$ in Krotov's method, it is necessary to define $J_{t}$
and formally minimize $J$, cf. Eq.~\eqref{eq:J}. We take~\cite{PRA.68.062308}
\begin{align} \label{eq:g}
  J_{t}\left[\left\{\Omega_{k}(t)\right\}\right]
  =
  \sum_{k} \frac{\lambda_{k}}{S(t)} \bigg[
    \Omega_{k}(t) - \Omega_{k}^{\rmref}(t)
  \bigg]^{2},
\end{align}
where $\Omega_{k}^{\rmref}(t)$ is a reference field for each $\Omega_{k}(t)$,
$S(t)$ the shape function from Eq.~\eqref{eq:S} and $\lambda_{k}$ a numerical
parameter that controls the magnitude of update in each iteration. By choosing
$\Omega_{k}^{\rmref}(t)$ to always be the respective field from the previous
iteration, $J_{t}$ will vanish as the optimization
converges~\cite{PRA.68.062308}. Hence, minimizing $J$ becomes identical to
minimizing $\error$, which is the important figure of merit that we seek to
minimize in the first place.

With $J_{t}$ from Eq.~\eqref{eq:g}, Krotov's method yields the update
equation~\cite{NJPhys.16.055012, PhD.reich}~\footnote{%
  Note the self-consistent nature of Eq.~\eqref{eq:update} where the update
  of the field $\Omega_{k}^{(i+1)}(t)$ at time $t$ on the left-hand side depends
  on the very same field and time on the right-hand side. In practice, this is
  solved by discretizing the time grid sufficiently fine such that for the
  update at time step $t=t_n$ on the left-hand side the corresponding values for
  fields (and states) at time step $t=t_{n-1}$ on the right-hand side are a good
  approximation. See Ref.~\cite{PRA.68.062308} for further details.
}
\begin{widetext}
  \begin{align} \label{eq:update}
    \Omega_{k}^{(i+1)}(t)
    &=
    \Omega_{k}^{(i)}(t)
    +
    \frac{S(t)}{\lambda_{k}} \rp\left\{%
      \sum_{l} \lol{%
        \op{\chi}^{(i)}_{l}(t)
        \Bigg|
        \frac{\partial \lop{L}(t)}{\partial
        \Omega_{k}} \Big|_{\{\Omega^{(i+1)}_{k'}(t)\}}
        \op{\rho}^{(i+1)}_{l}(t)
      }
    \right\}.
  \end{align}
\end{widetext}
$\{\op{\rho}^{(i+1)}_{l}(t)\}$ are forward propagated initial states
$\{\op{\rho}_{l}(0)\}$, i.e., solutions to the Lindblad master equation
\begin{equation} \label{eq:fw_states}
  \frac{\dd}{\dd t} \op{\rho}^{(i+1)}_{l}(t)
  =
  \lop{L}^{(i+1)}(t) \left[\op{\rho}^{(i+1)}_{l}(t)\right]
\end{equation}
under the \emph{new} fields $\{\Omega_{k}^{(i+1)}(t)\}$.
$\{\op{\chi}^{(i)}_{l}(t)\}$ are so-called co-states, which are solutions to the
adjoint equation of motion
\begin{equation} \label{eq:co_states}
  \frac{\dd}{\dd t} \op{\chi}^{(i)}_{l}(t)
  =
  \lop{L}^{(i)\dagger}(t) \left[\op{\chi}^{(i)}_{l}(t)\right]
\end{equation}
under the \emph{old} fields $\{\Omega_{k}^{(i)}(t)\}$ with boundary condition
\begin{align}
  \op{\chi}^{(i)}_{l}(T)
  =
  - \frac{\partial \error}{\partial \op{\rho}_{l}(T)}
  \Big|_{\{\op{\rho}^{(i)}_{l'}(T)\}}.
\end{align}
The initial states $\{\op{\rho}_{l}(0)\}$ are in our case given by the four
logical basis states from Eq.~\eqref{eq:fockL} or Eq.~\eqref{eq:binomL}
depending on which encoding we want to optimize for by minimizing
Eq.~\eqref{eq:JT}.

We use the QDYN library~\cite{qdyn} for solving all equations of motion and for
Krotov's method. The NLopt library~\cite{nlopt} is used for the gradient-free
optimizations.

\begin{table}[t!]
  \centering
  \caption{%
    Parameters for cavities A and B and transmon C, taken from
    Ref.~\cite{PRX.8.021073}.
  }
  \begin{tabular*}{\linewidth}{l@{\extracolsep{\fill}}cr}
    \hline
    frequency cavity A & $\omega_{\rma}/2\pi$ & \SI{5.554}{\giga\hertz}
    \\
    frequency cavity B & $\omega_{\rmb}/2\pi$ & \SI{6.543}{\giga\hertz}
    \\
    base frequency transmon C & $\omega_{\rmc}/2\pi$ & \SI{5.901}{\giga\hertz}
    \\
    anharmonicity transmon C & $\alpha_{\rmc}/2\pi$ & \SI{74}{\mega\hertz}
    \\
    coupling between A and C & $g_{\rma}/2\pi$ & \SI{-19.921}{\mega\hertz}
    \\
    coupling between B and C & $g_{\rmb}/2\pi$ & \SI{28.417}{\mega\hertz}
    \\ \hline
    amplitude driving field 1 & $\Omega_{1}/2\pi$ & \SI{94.200}{\mega\hertz}
    \\
    amplitude driving field 2 & $\Omega_{2}/2\pi$ & \SI{229.725}{\mega\hertz}
    \\
    frequency driving field 1 & $\omega_{1}/2\pi$ & \SI{6058.000}{\mega\hertz}
    \\
    frequency driving field 2 & $\omega_{2}/2\pi$ & \SI{7049.624}{\mega\hertz}
    \\
    ramping time & $\tau$ & \SI{50}{\nano\second}
    \\ \hline
    relaxation of cavity A & $T_{1}^{\rma}$ & \SI{1000}{\micro\second}
    \\
    relaxation of cavity B & $T_{1}^{\rmb}$ & \SI{1000}{\micro\second}
    \\
    relaxation of transmon C & $T_{1}^{\rmc}$ & \SI{50}{\micro\second}
    \\
    dephasing of cavity A & $T_{\phi}^{\rma}$ & $\infty$ 
    \\
    dephasing of cavity B & $T_{\phi}^{\rmb}$ & $\infty$ 
    \\
    dephasing of transmon C & $T_{\phi}^{\rmc}$ & \SI{50}{\micro\second}
    \\ \hline
  \end{tabular*}
  \label{tab:params}
\end{table}

\begin{figure}[tb!]
  \centering
  \includegraphics{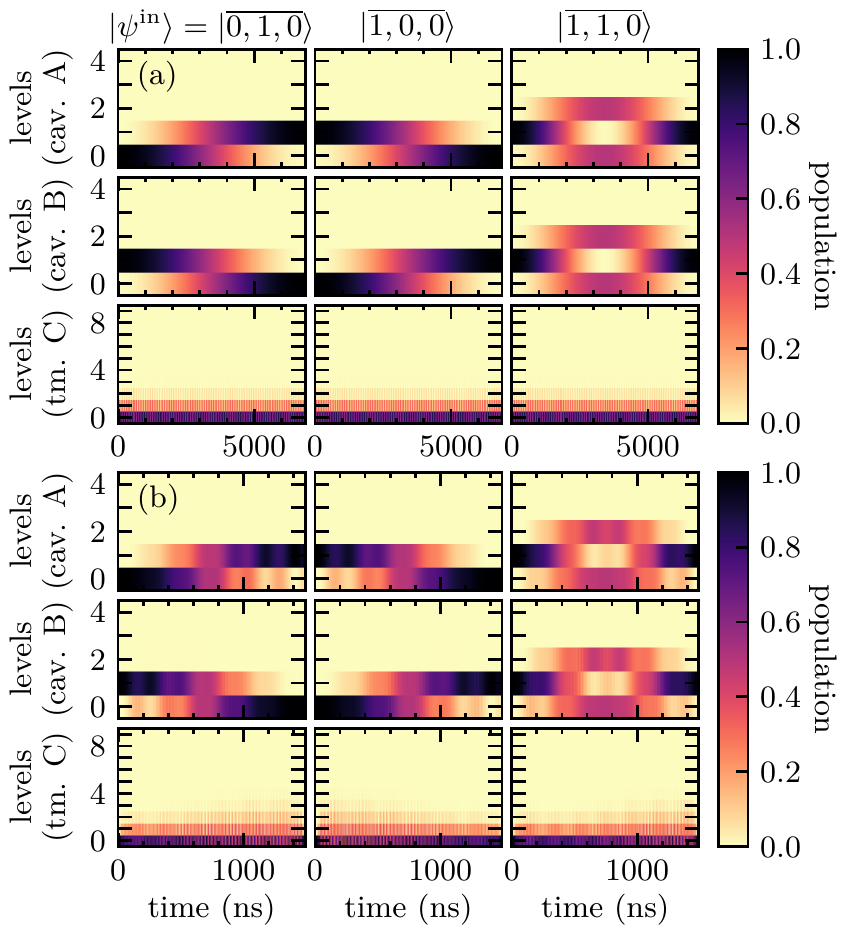}
  \caption{%
    Population dynamics for the three initial states $\dket{0,1,0},
    \dket{1,0,0}$ and $\dket{1,1,0}$ with $\dket{0,0,0}$ omitted as it does not
    give rise to any cavity excitations. (a) Dynamics under the original
    two-tone protocol of Ref.~\cite{PRX.8.021073} using the parameters from
    Table~\ref{tab:params}. (b) Dynamics as in (a) but for a three-tone
    protocol with optimized but constant amplitude $\Omega_{3}$ and frequency
    $\omega_{3}$. Note the different time scales. See main text for details.
    The term ''levels'' refers to the bare Fock states $\ket{\na}$ and
    $\ket{\nb}$ in case of cavities A and B and to the anharmonic ladder states
    $\ket{\nc}$ of the bare transmon in case of transmon C.
  }
  \label{fig:pop}
\end{figure}

\section{Engineering Strong Beamsplitter Interaction via Sideband Transitions}
\label{sec:strong}
In this section, we demonstrate how to use OCT in order to engineer
a beamsplitter interaction between cavities A and B that is stronger than the
one presented in Ref.~\cite{PRX.8.021073}.

\subsection{Two-tone vs.\ three-tone protocol}
\label{subsec:2vs3}
We take the physical parameters as reported in Ref.~\cite{PRX.8.021073}, cf.
Table~\ref{tab:params}, and start by analyzing the original two-tone protocol.
The two tones' amplitudes, $\Omega_{1}$ and $\Omega_{2}$, and frequencies,
$\omega_{1}$ and $\omega_{2}$, are chosen to satisfy the resonance
condition~\eqref{eq:reson} and therefore give rise to the desired beamsplitter
interaction, cf. Eq.~\eqref{eq:HabBS}. If we assume Fock encoding in
Eq.~\eqref{eq:JT}, the subspace for which to test the protocol is defined by the
four initial states $\dket{0,0,0}, \dket{0,1,0}, \dket{1,0,0}$ and
$\dket{1,1,0}$. Figure~\ref{fig:pop}(a) shows the population dynamics for these
initial states under the original two-tone protocol~\cite{PRX.8.021073}. As
expected, the dynamics swaps the initial states of cavity A and B for
$\dket{0,1,0}$ and $\dket{1,0,0}$ and leaves the states $\dket{0,0,0}$ and
$\dket{1,1,0}$ invariant at final time $T=\SI{6780}{\nano\second}$. This
invariance does not hold at intermediate times, cf.\ the dynamics of
$\dket{1,1,0}$. Note that the transmon is only weakly excited (its time-averaged
ground state population is 0.73, cf.  Eq.~\eqref{eq:avg_pop_tm_0}), which is in
agreement with the theory of the two-tone protocol~\cite{PRA.99.012314}.The two
tones are switched on and off smoothly by a ramp $S(t)$. This transfers the
transmon smoothly from its ground state into an energetically low-lying Floquet
state at intermediate times and back to the ground state at final time. This is
the reason for the small but non-vanishing population in some lower bare
transmon levels seen in Fig.~\ref{fig:pop}(a).

We now add a third control field with amplitude $\Omega_{3}$ and frequency
$\omega_{3}$ with the purpose to realize the desired swap, cf.
Eq.~\eqref{eq:dmap_trgt}, in a shorter total time $T$. As outlined in
Sec.~\ref{subsec:oct}, we use a gradient-free optimization to find optimized
values for the three parameters, i.e., $\Omega_{3}, \omega_{3}, T$. We kept the
parameters of the other two control fields, $\Omega_{1}, \Omega_{2}, \omega_{1},
\omega_{2}$, fixed to limit the number of optimization parameters and thus ease
the optimization procedure. We find the optimized parameters for the third drive
to be
\begin{align} \label{eq:3rdtone}
  \Omega_{3}/2\pi
  =
  \SI{271.093}{\mega\hertz},
  \qquad
  \omega_{3}/2\pi
  =
  \SI{6.749}{\giga\hertz}
\end{align}
and the protocol duration $T = \SI{1492}{\nano\second}$, which is about five
times shorter than $T = \SI{6780}{\nano\second}$ in case of the two-tone
protocol. Figure~\ref{fig:pop}(b) shows the corresponding population dynamics
for the three-tone protocol. In comparison with the dynamics of the original
two-tone protocol, cf. Fig.~\ref{fig:pop}(a), the dynamics of the three-tone
protocol looks very similar but is approximately five times faster. We find the
coherent errors to be $\error_{2} = 0.5\%$ for the two-tone protocol and
$\error_{3} = 2.6\%$ for the three-tone protocol. When decoherence is taken into
account, the errors increase to $\error_{2} = 8.6\%$ and $\error_{3} = 5.0\%$
for the two-tone and three-tone protocol, respectively. As expected, the
increase in error is much smaller for the significantly faster three-tone
protocol compared to the original two-tone protocol and compensates the
previously larger coherent error of the three-tone protocol. We will analyze
the impact of decoherence in more detail in Sec.~\ref{sec:diss}.

\begin{figure}[tb!]
  \centering
  \includegraphics{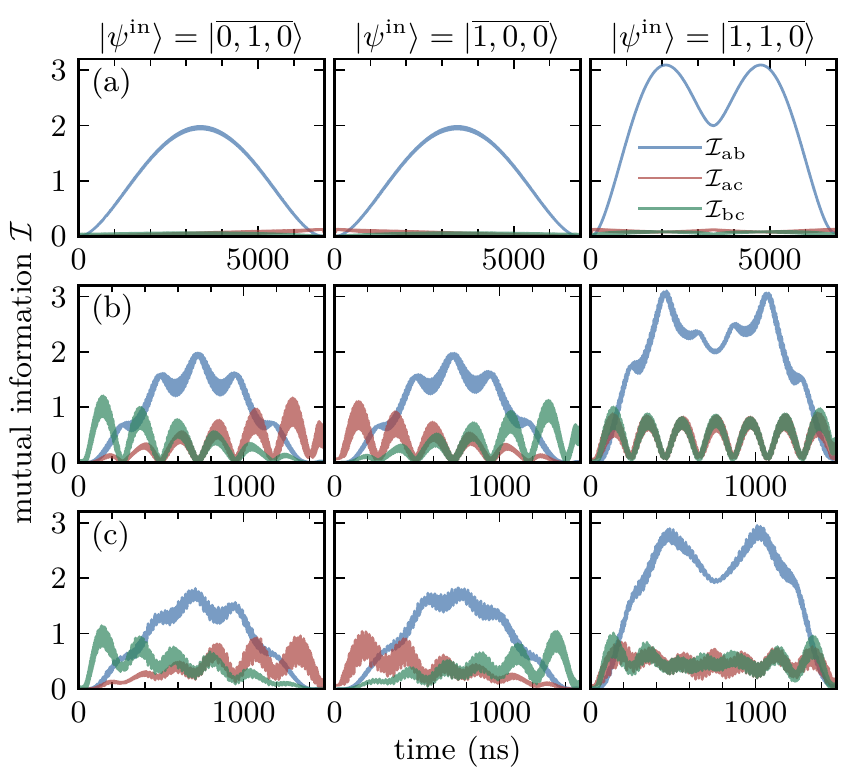}
  \caption{%
    Mutual information $\mathcal{I}$ between all pairs of subsystems A, B and C.
    Panels (a) and (b) correspond to the two- and three-tone protocol and the
    population dynamics shown in Fig.~\ref{fig:pop}(a) and (b), respectively.
    Panel (c) corresponds to the three-tone protocol that has been further
    fine-tuned by a gradient-based optimization technique (see main text for
    details).
  }
  \label{fig:corr}
\end{figure}

In order to understand the similarities and differences of the two- and
three-tone protocols, we inspect the correlations between any two of the three
subsystems of cavities A and B and transmon C as a function of time.
Figure~\ref{fig:corr}(a) and (b) show the mutual information $\mathcal{I}$, as
a measure for the correlations between two subsystems~\cite{Henderson2001}, for
the two- and three-tone protocol, respectively. The mutual information between
two subsystems, say cavities A and B, is defined as
\begin{align}
  \mathcal{I}_{\mathrm{ab}}
  =
  \mathcal{P}(\op{\rho}_{\rma})
  +
  \mathcal{P}(\op{\rho}_{\rmb})
  -
  \mathcal{P}(\op{\rho}_{\mathrm{ab}}),
\end{align}
with $\mathcal{P}(\op{\rho}_{\rmx}) = - \tr\{\op{\rho}_{\rmx}
\ln(\op{\rho}_{\rmx})\}$ the von Neumann entropy of state $\op{\rho}_{\rmx}$ and
$\op{\rho}_{\mathrm{ab}} = \tr_{\rmc}\{\op{\rho}\}$, $\op{\rho}_{\rma}
= \tr_{\rmb}\{\op{\rho}_{\mathrm{ab}}\}$ and $\op{\rho}_{\rmb}
= \tr_{\rma}\{\op{\rho}_{\mathrm{ab}}\}$ the reduced states of subsystems AB,
A and B, respectively, calculated from the state $\op{\rho} = \op{\rho}(t)$ of
the full tripartite system. As can be seen for the two-tone protocol, cf.
Fig.~\ref{fig:corr}(a), only the two cavities build up correlations over time,
whereas the transmon C stays uncorrelated with both at all times. At final time
$T$, both cavities are again uncorrelated. This behavior changes for the
three-tone protocol, cf.  Fig.~\ref{fig:corr}(b), as it gives rise to additional
intermediate correlations between both cavities and the transmon as well as
remaining, non-vanishing correlations at final time $T$.  Closer inspection of
the dynamics for the initial state $\dket{1,0,0}$ ($\dket{0,1,0}$) reveals that,
in the first half of the protocol, cavity A (B) primarily correlates with the
transmon while in the second half primarily cavity B (A) correlates with the
transmon. In particular correlations that are built up in the second half do not
vanish at final time $T$. This is a reason for the larger coherent error
$\error_{3}$ of the three-tone protocol. We show in
Appendix~\ref{sec:app:construct} that, despite the emerging correlations between
cavities and transmon, the three-tone protocol still engineers the intended
beamsplitter interaction~\eqref{eq:HabBS}.

\begin{figure*}[tb!]
  \centering
  \includegraphics{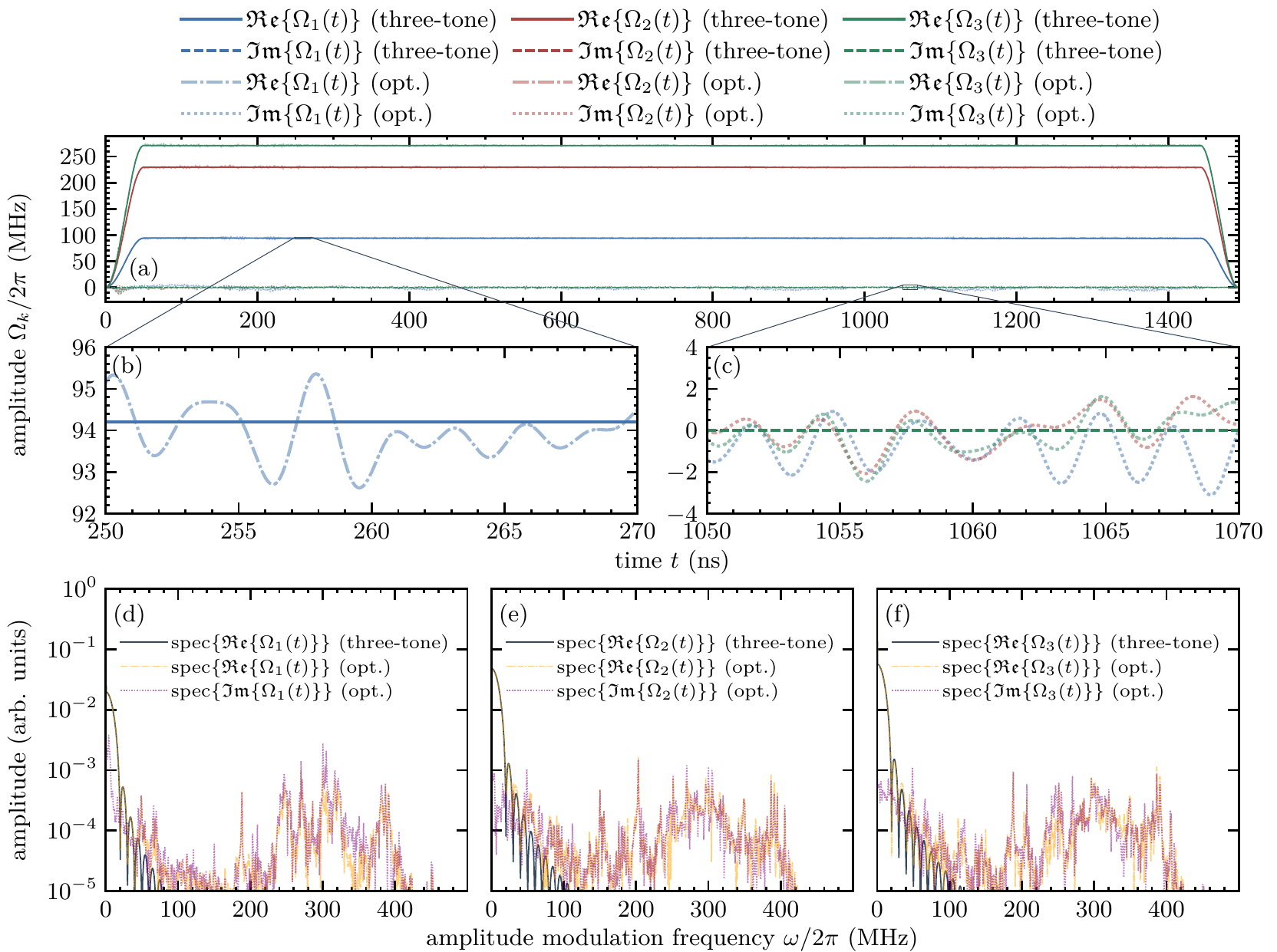}
  \caption{%
    Panel (a) shows the time-dependent amplitudes $\Omega_{k}(t)$ of the
    three-tone protocol (solid and dashed lines) and their fine-tuned versions
    (dashed-dotted and dotted lines), while (b) and (c) show a close-up view.
    The coherent errors for the three-tone protocol and its fine-tuned version
    are $\error_{3} = 2.6\%$ and $\error_{3,\rmgrad} = 0.2\%$, respectively.
    Panels (d)-(f) show the spectra (the absolute value of the Fourier
    transform) of $\rp\{\Omega_{k}(t)\}$ and $\ip\{\Omega_{k}(t)\}$ for the
    three-tone protocol (solid lines) and its fine-tuned version (dashed-dotted
    and dotted lines), respectively. Note that $\ip\{\Omega_{k}(t)\}$ is zero
    for the three-tone protocol and hence its spectrum is omitted. The
    dashed-dotted and dotted lines are heavily overlapping and are thus hard to
    distinguish visually.
    The base-frequencies for the three tones are given in Table~\ref{tab:params}
    and Eq.~\eqref{eq:3rdtone}.
  }
  \label{fig:fields}
\end{figure*}

\subsection{Fine-tuned three-tone protocol}
\label{subsec:grad}
A possibility to reduce the coherent error $\error_{3}$ of the three-tone
protocol is to fine-tune it further with gradient-based optimization, as
outlined in Sec.~\ref{subsec:oct}, using the constant values of the three-tone
protocol (up to ramping times) as guess fields. While each field $\Omega_{k}(t)$
has its base frequency $\omega_{k}$, cf. Eq.~\eqref{eq:ham}, allowing for fully
time-dependent and complex $\Omega_{k}(t)$ introduces new frequencies, i.e., the
protocol is strictly speaking no longer a three-tone protocol. In order to
prevent the bandwidth of the amplitude modulations to become too large and
experimentally unfeasible, we truncate the spectrum of $\rp\{\Omega_{k}(t)\}$
and $\ip\{\Omega_{k}(t)\}$ after each iteration of the optimization by
multiplying the spectrum with the spectral shape function
\begin{align}
  S_{\mathrm{bw}}(\omega)
  =
  \begin{cases}
    1,
    &\qquad |\omega| \in [0, \omega_{\mathrm{co}} - \Delta \omega_{\mathrm{co}}]
    \\
    \sin^{2} \left(
      \pi \frac{|\omega| - \omega_{\mathrm{co}}}{2 \Delta \omega_{\mathrm{co}}}
    \right),
    &\qquad |\omega| \in (\omega_{\mathrm{co}} - \Delta \omega_{\mathrm{co}},
    \omega_{\mathrm{co}}],
    \\
    0,
    &\qquad \mathrm{else},
  \end{cases}
\end{align}
where $\omega_{\mathrm{co}}$ is a cutoff frequency. Here, $\Delta
\omega_{\mathrm{co}}$ is necessary to truncate the spectra smoothly and
guarantee a smooth shape of $\Omega_{k}(t)$ in time domain.

Figure~\ref{fig:fields}(a)-(c) compares the real and imaginary parts of
$\Omega_{k}(t)$ for the three-tone protocol with the further fine-tuned version.
As can be seen, the gradient-based optimization adapts the amplitudes slightly
by adding minor oscillations. Despite these apparently small differences, the
coherent protocol error reduces from $\error_{3} = 2.6\%$ to $\error_{3,\rmgrad}
= 0.2\%$. The spectra of $\rp\{\Omega_{k}(t)\}$ and $\ip\{\Omega_{k}(t)\}$ of
the three-tone protocol and its further fine-tuned version are shown in
Fig.~\ref{fig:fields}(d)-(f). The three-tone protocol has a single dominant peak
at modulation frequency $\omega = 0$ with only minor non-zero elements due to
the ramping. After optimization, new amplitude modulation frequencies up to
$\omega/2\pi \sim \SI{400}{\mega\hertz}$ appear, reflecting our choice of
$\omega_{\mathrm{co}}/2\pi = \SI{500}{\mega\hertz}$ and $\Delta
\omega_{\mathrm{co}}/2\pi = \SI{100}{\mega\hertz}$ for truncating the spectra.
Compared to the spectral amplitude of the central peak at $\omega = 0$, these
new frequencies have spectral amplitudes that are at least two orders of
magnitude smaller. This is consistent with the amplitudes $\Omega_{k}(t)$
remaining almost constant in time with only small oscillations on top, cf.
Fig.~\ref{fig:fields}(b) and (c). The effect of the modulations can be seen in
Fig.~\ref{fig:corr}(c), which shows the mutual information between the three
subsystems. While the overall structure of the correlation dynamics is preserved
compared to the three-tone protocol in Fig.~\ref{fig:corr}(b), the fine-tuned
amplitudes $\Omega_{k}(t)$ erase all correlations at final time $T$. This
concerns especially those correlations between cavity A/B and the transmon
C built up when starting in states $\dket{0,1,0}$ or $\dket{1,0,0}$ which do not
vanish at final time $T$ under the non-fine-tuned three-tone protocol. Despite
the difference in the correlation dynamics and final errors, the population
dynamics for the fine-tuned protocol is visually almost identical to the
three-tone protocol shown in Fig.~\ref{fig:pop}(b) (data not shown).

Note that technically, it would also be possible to carry out the optimization
with a single field instead of optimizing the three tones individually.
Motivated by Eq.~\eqref{eq:ham'}, one could for instance define the effective
field $\pulse_{\rmeff}(t) = \sum_{k} \Omega_{k}(t) e^{- \im \delta_{k} t}$ and
optimize its real and imaginary part. This carries the same information.
However, an optimization with three individual tones allows for more flexibility
when controlling each field's update, e.g.\ by truncating the spectra (as used
above) or by choosing which tones should be updated at all.

It is of course possible to apply the gradient-based optimization --- including
restricting the amplitude modulation frequencies by spectral truncation --- also
to the two-tone protocol directly. This lowers the coherent error from
$\error_{2} = 0.5\%$ to $\error_{2,\rmgrad} < 0.1\%$. The changes to the
amplitudes $\Omega_{k}(t)$ are even smaller than the ones shown in
Fig.~\ref{fig:fields} for the three-tone protocol. However, for both the two-
and three-tone protocol, decoherence is the dominant source of error which
increases to $\error_{2,\rmgrad} = 9.2\%$ and $\error_{3,\rmgrad}
= 3.2\%$, respectively, once decoherence is taken into account.

A subtle but important fact can be noticed when comparing the change between the
coherent protocol errors under the fine-tuned protocols, once decoherence
is accounted for. In detail, the increase due to decoherence is slightly larger
for the fine-tuned versions of the two- and three-tone protocols
compared to their non-fine-tuned versions despite unchanged duration
$T$. This is due to the fact that --- for reasons to keep the numerical costs
manageable --- the optimization itself is carried out entirely in Hilbert space,
i.e., without taking decoherence into account \emph{explicitly}. Instead, we
account for it \emph{implicitly} by penalizing control solutions that involve
excitation of higher transmon levels which suffer more from decoherence.
Although the dynamics under the fine-tuned protocols does not utilize higher
transmon excitations, it exploits coherences between energetically low-lying but
populated levels --- especially those between the transmon's bare ground and
first excited state. Hence, any deviation from the desired dynamics of the
coherences due to decoherence causes the protocol error to increase. In case of
the two-tone protocol, the increase for the fine-tuned version is larger than
that of the original version, illustrating the fine-tuned protocol's somewhat
increased sensitivity to decoherence.

\begin{table}[t!]
  \centering
  \caption{%
    Protocol errors for the original two-tone protocol, $\error_{2}$, the
    three-tone protocol, $\error_{3}$, and their respective fine-tuned
    versions, $\error_{2,\rmgrad}$ and $\error_{3,\rmgrad}$. The errors are
    given for both Fock and binomial encoding, cf. Eq.~\eqref{eq:JT}.
    Decoherence times are as in Table~\ref{tab:params}.
  }
  \begin{tabular*}{\linewidth}{c@{\extracolsep{\fill}}ccc}
    \hline
    & coherent & error including & encoding
    \\
    & error & decoherence &
    \\
    \hline
    $\error_{2}$                          &  $0.5\%$ &  $8.6\%$ & Fock
    \\
    $\error_{2,\rmgrad}$                  & $<0.1\%$ &  $9.2\%$ & Fock
    \\
    $\error_{3}$                          &  $2.6\%$ &  $5.0\%$ & Fock
    \\
    $\error_{3,\rmgrad}$                  &  $0.2\%$ &  $3.2\%$ & Fock
    \\
    \hline
    $\error^{\mathrm{binom}}_{2}$         & $43.9\%$ & $53.9\%$ & binomial
    \\
    $\error^{\mathrm{binom}}_{2,\rmgrad}$ &  $0.8\%$ & $35.3\%$ & binomial
    \\
    $\error^{\mathrm{binom}}_{3}$         & $58.7\%$ & $60.6\%$ & binomial
    \\
    $\error^{\mathrm{binom}}_{3,\rmgrad}$ &  $2.0\%$ & $11.0\%$ & binomial
    \\
    \hline
  \end{tabular*}
  \label{tab:errors}
\end{table}

So far we have discussed whether the engineered dynamics behaves as intended
when the qubit is encoded in the two lowest Fock states of each cavity.
However, as emphasized earlier, it would be advantageous to have a protocol that
works in a codeword-agnostic way. Thus, as an alternative to the qubit being
encoded in the two lowest Fock states, we also employ a binomial
encoding~\cite{PRX.6.031006}. In this case, the error --- in the following
called $\error^{\mathrm{binom}}$ --- is still given by Eq.~\eqref{eq:JT} but the
latter is evaluated for the logical basis states of Eq.~\eqref{eq:binomL}. We
find coherent errors $\error^{\mathrm{binom}}_{2} = 43.9\%$ and
$\error^{\mathrm{binom}}_{3} = 58.7\%$ for the two-tone protocol of
Ref.~\cite{PRX.8.021073} and the three-tone protocol, respectively. Taking these
protocols again as starting point for a gradient-based optimization --- here
without frequency truncation --- we find coherent errors of
$\error^{\mathrm{binom}}_{2,\rmgrad} = 0.8\%$ and
$\error^{\mathrm{binom}}_{3,\rmgrad} = 2.0\%$, respectively~\footnote{The
optimizations have been carried out without frequency truncation in order to
keep the number of iterations sufficiently small}. While the two- and three-tone
protocol do not act as codeword-agnostic as desired, our results indicate that
it is possible to adapt each protocol for a given encoding using OCT.  Note that
these errors are obtained without taking decoherence in account. With
decoherence, they become $\error^{\mathrm{binom}}_{2} = 53.9\%$ and
$\error^{\mathrm{binom}}_{3} = 60.6\%$ for the two- and three-tone protocol,
respectively, and $\error^{\mathrm{binom}}_{2,\rmgrad} = 35.3\%$ and
$\error^{\mathrm{binom}}_{3,\rmgrad} = 11.0\%$ for the fine-tuned versions.
A summary of all errors, with and without decoherence, is provided in
Table~\ref{tab:errors}.

\subsection{Analysis of the beamsplitter interaction in the three-tone protocol}
\label{subsec:ana}
We now seek to understand why a third tone gives rise to significantly faster
swaps, respectively stronger beamsplitter interaction. To this end, we first
notice that the gradient-free optimization chooses the third frequency
$\omega_{3}$, cf. Eq.~\eqref{eq:3rdtone}, such as to give rise to near-resonant
sideband couplings between cavity A/B and transmon C. The sideband couplings are
induced by the beating between the third drive and the first two drives. While
satisfying Eq.~\eqref{eq:reson} activates the beamsplitter interaction in the
original two-tone protocol, we can define a similar resonance condition that
needs to be fulfilled for activating the cavity-transmon sideband couplings. It
reads
\begin{align} \label{eq:reson_sb}
  \som_{\rma} - \som_{\rmc}
  &=
  \omega_{3} - \omega_{2} + \Delta_{\rma},
  \notag \\
  \som_{\rmb} - \som_{\rmc}
  &=
  \omega_{3} - \omega_{1} + \Delta_{\rmb},
\end{align}
where $\som_{\rmc}$ is the Stark-shifted version of $\omega_{\rmc}$ and
$\Delta_{\rma}$ and $\Delta_{\rmb}$ are detunings from the corresponding perfect
sideband couplings between cavity A/B and transmon C, respectively. In order to
fulfill the beamsplitter resonance condition~\eqref{eq:reson}, we need
$\Delta_{\rma} \approx \Delta_{\rmb}$. We thus set $\Delta_{\rma}
= \Delta_{\rmb} = \Delta$.

In the following, we use a similar approach as in Ref.~\cite{PRA.99.012314},
where the effective beamsplitter Hamiltonian~\eqref{eq:HabBS} was derived
analytically from the tripartite system, including the transmon. The derivation
just assumed a two-tone protocol with both frequencies fulfilling the resonance
condition~\eqref{eq:reson}. Here, we modify the approach of
Ref.~\cite{PRA.99.012314} to include the sideband couplings. We also seek to
derive an effective Hamiltonian that describes the effective interaction of the
two cavities. In contrast to the two-tone protocol, our derivation needs to
capture both the cavity-cavity beamsplitter interaction, generated by fulfilling
the resonance condition~\eqref{eq:reson}, as well as the cavity-transmon
sideband coupling, generated by fulfilling Eq.~\eqref{eq:reson_sb}
near-resonantly, i.e., with a small, but non-zero $\Delta$. Our derivation can
thus be seen as an extension of the derivation done in
Ref.~\cite{PRA.99.012314}. Ultimately, we will compare SWAP times predicted
analytically by our derivation (carried out in the following) and
semi-analytically using a method from Ref.~\cite{PRA.99.012314} with
numerically obtained ones.

First, we assume a weak transmon anharmonicity $\alpha_{\rmc} \ll
|\omega_{\rma,\rmb} - \omega_{\rmc}|$ and diagonalize the quadratic and
field-free part of Hamiltonian~\eqref{eq:ham} to obtain the eigenmodes. The
associated lowering operators for the eigenmodes are $\op{A}, \op{B}$ and
$\op{C}$. For weak transmon-cavity couplings, one can identify $\op{A}$ and
$\op{B}$ as the ``cavity-like'' eigenmodes that have the largest overlap with
the bare cavity modes $\op{a}$ and $\op{b}$ and $\op{C}$ is a ``transmon-like''
eigenmode. Next, we express the bare transmon operator $\op{c}$ as a function of
the eigenmodes of the quadratic, field-free Hamiltonian. In addition to the
coupling-induced mode mixing of the bare transmon modes, we also want to capture
the effect of the drives and express it in the eigenmode representation of
$\op{c}$. To this end, we exploit that for weakly anharmonic transmons the major
effect of the drives is to induce a linear displacement of the transmon mode.
Combining the coupling-induced mode mixing and the drive-induced displacement of
the mode, we find
\begin{align} \label{eq:c}
  \op{c}
  \rightarrow
  \xi_{\rma} \op{A}
  +
  \xi_{\rmb} \op{B}
  +
  \xi_{\rmc} \op{C}
  +
  \sum_{k=1}^{3} \xi_{k} e^{- \im \omega_{k} t},
\end{align}
where $\xi_{\rma (\rmb)} \approx g_{\rma (\rmb)}/\delta_{\rma (\rmb)}$,
$\xi_{\rmc} \approx 1$, and $\xi_{k} = \Omega_{k}/\delta_{k}$ for $k=1,2,3$. By
substituting Eq.~\eqref{eq:c} into Hamiltonian~\eqref{eq:ham} and transforming
into a rotating frame --- similar to that from Hamiltonian~\eqref{eq:ham} to
Hamiltonian~\eqref{eq:ham'} --- we arrive at
\begin{align} \label{eq:ham_eff_1}
  \op{H}_{\rmeff,1}(t)
  &=
  g_{\rmAB} \op{A}^{\dagger} \op{B}
  +
  g_{\rmAB}^{*} \op{A} \op{B}^{\dagger}
  \notag \\
  &\quad
  +
  g_{\rma}^{\rmsb} e^{\im \Delta t} \op{A}^{\dagger} \op{\sigma}_{-}
  +
  g_{\rma}^{\rmsb*} e^{- \im \Delta t} \op{A} \op{\sigma}_{+}
  \notag \\
  &\quad
  +
  g_{\rmb}^{\rmsb} e^{\im \Delta t} \op{B}^{\dagger} \op{\sigma}_{-}
  +
  g_{\rmb}^{\rmsb*} e^{- \im \Delta t} \op{B} \op{\sigma}_{+},
\end{align}
where we have only kept resonant and near-resonant terms. We also truncate the
transmon Hilbert space to two levels, replacing $\op{C}$ and $\op{C}^{\dagger}$
by $\op{\sigma}_{-}$ and $\op{\sigma}_{+}$. The cavity-cavity coupling
$g_{\rmAB}$ and transmon-cavity sideband couplings $g_{\rma}^{\rmsb}$ and
$g_{\rmb}^{\rmsb}$ are given by
\begin{align} \label{eq:sb_ints}
  g_{\rmAB}
  &=
  - 2 \alpha_{\rmc} \xi_{1} \xi_{2}^{*} \xi_{\rma}^{*} \xi_{\rmb},
  \notag \\
  g_{\rma}^{\rmsb}
  &=
  - 2 \alpha_{\rmc} \xi_{2}^{*} \xi_{3} \xi_{\rma}^{*} \xi_{\rmc},
  \notag \\
  g_{\rmb}^{\rmsb}
  &=
  - 2 \alpha_{\rmc} \xi_{1}^{*} \xi_{3} \xi_{\rmb}^{*} \xi_{\rmc}.
\end{align}
In a further transformation, we move into a rotating frame with respect to
$\Delta$ where Eq.~\eqref{eq:ham_eff_1} becomes time-independent,
\begin{align} \label{eq:ham_eff_2}
  \op{H}_{\rmeff,2}
  &=
  \frac{\Delta}{2} \op{\sigma}_{\rmz}
  +
  g_{\rmAB} \op{A}^{\dagger} \op{B}
  +
  g_{\rmAB}^{*} \op{A} \op{B}^{\dagger}
  \notag \\
  &\quad
  +
  g_{\rma}^{\rmsb} \op{A}^{\dagger} \op{\sigma}_{-}
  +
  g_{\rma}^{\rmsb*} \op{A} \op{\sigma}_{+}
  \notag \\
  &\quad
  +
  g_{\rmb}^{\rmsb} \op{B}^{\dagger} \op{\sigma}_{-}
  +
  g_{\rmb}^{\rmsb*} \op{B} \op{\sigma}_{+}.
\end{align}
This Hamiltonian describes the dynamics of two degenerate modes coupled to each
other and to a detuned two-level system. In the limit when $|g_{\rma}^{\rmsb}|,
|g_{\rmb}^{\rmsb}| \ll |\Delta|$ holds, Eq.~\eqref{eq:ham_eff_2} can be
diagonalized perturbatively in $|g_{\rma,\rmb}^{\rmsb} / \Delta|$. To second
order, we obtain
\begin{subequations}
\begin{align} \label{eq:ham_eff_3}
  \op{H}_{\rmeff,3}
  =
  \op{g}_{\rmBS} \op{A}^{\dagger} \op{B}
  +
  \op{g}_{\rmBS}^{*} \op{A} \op{B}^{\dagger}
\end{align}
with
\begin{align} \label{eq:gBS}
  \op{g}_{\rmBS}
  =
  g_{\rmAB} \uop - \frac{g_{\rma}^{\rmsb} g_{\rmb}^{\rmsb*}}{\Delta}
  \op{\sigma}_{\rmz}.
\end{align}
\end{subequations}
Note that we neglect terms $\sim \op{A}^{\dagger} \op{A}$ and $\sim
\op{B}^{\dagger} \op{B}$ since they represent shifts of the cavity eigenmodes
and can be simply compensated by adapting the drive frequencies.
Equation~\eqref{eq:ham_eff_3} resembles the desired beamsplitter interaction,
cf. Eq.~\eqref{eq:HabBS}, but has a slightly more complex structure of the
operator $\op{g}_{\rmBS}$. It contains the ``standard'' beamsplitter interaction
--- here given by the term scaling with $g_{\rmAB}$ --- that originates from
having two drives which fulfill the resonance condition~\eqref{eq:reson} and
additionally the sideband-induced couplings between cavities and transmon. The
latter originate from the third drive and effectively amplify the beamsplitter
interaction. This gives rise to the faster SWAP operation as observed in
Fig.~\ref{fig:pop}.

\begin{figure}[tb!]
  \centering
  \includegraphics[width=86mm]{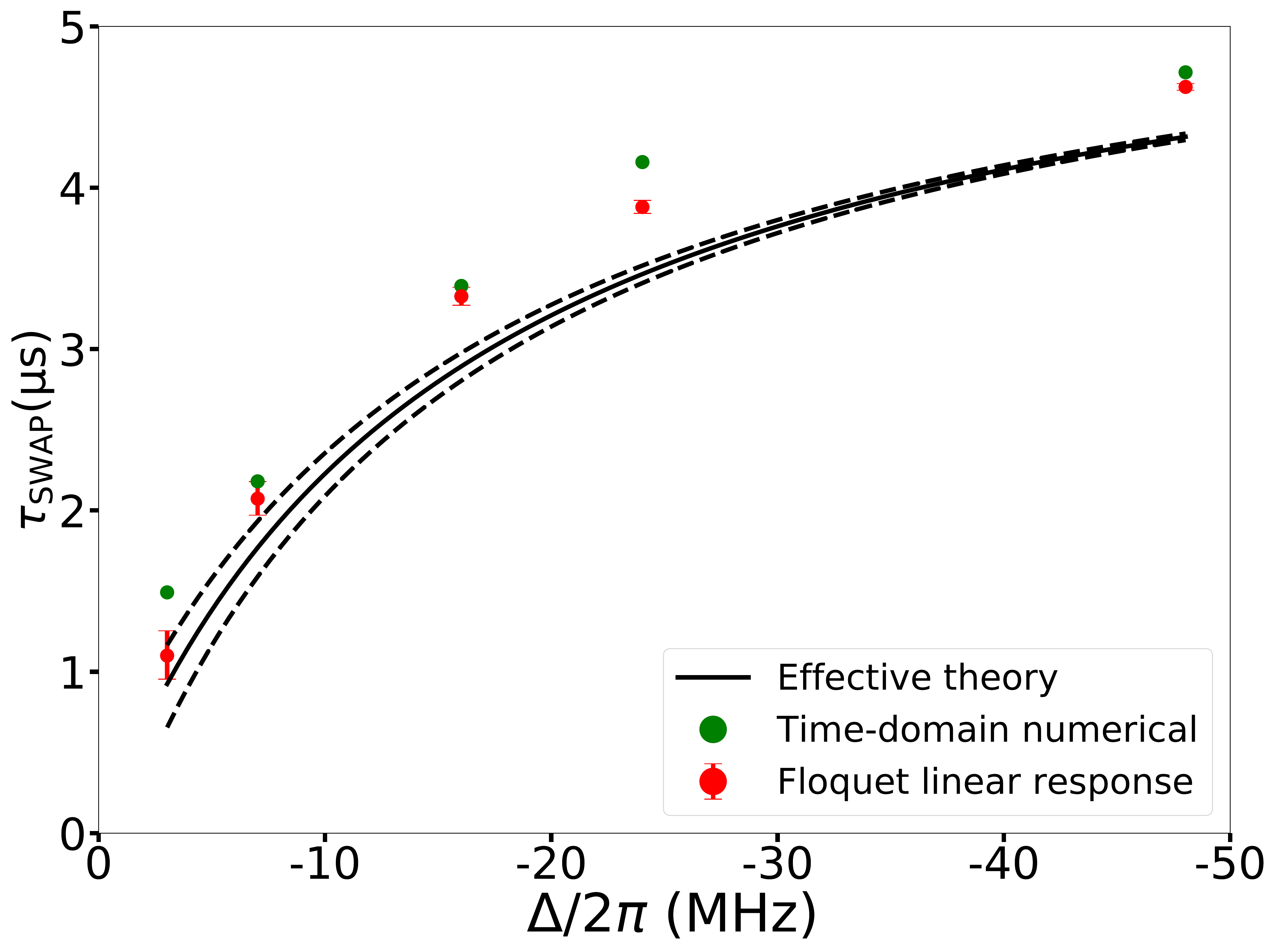}
  \caption{%
    Comparison of the protocol times as obtained numerically (green dots) and
    from analytical calculations including one from Floquet theory in the spirit
    of Ref.~\cite{PRA.99.012314} (red dots) and one using the effective theory
    from Eqs.~\eqref{eq:ham_eff_3} and~\eqref{eq:gBS} (black solid line).
    Note that approximated versions of $\Delta$ and other Stark-shifted
    parameters have been used here as the exact Stark shifts are not known
    analytically. This is reflected by the error bars for the Floquet theory
    and the black dashed lines for the effective theory. The lower (upper) error
    bars are obtained using a value of $\Delta/2\pi$ that is $-(+)
    \SI{1}{\mega\hertz}$ away from the value indicated on the horizontal axis.
  }
  \label{fig:SWAP_times}
\end{figure}

\begin{table}[t!]
  \centering
  \caption{%
    Further three-tone protocols, obtained via gradient-free optimization, with
    all other parameters as given in Table.~\ref{tab:params} and $\Omega_{3}$ as
    in Eq.~\eqref{eq:3rdtone}.
  }
  \begin{tabular*}{\linewidth}{c@{\extracolsep{\fill}}ccc}
    \hline
    $\omega_{2}/2\pi$~(MHz) & $\omega_{3}/2\pi$~(MHz) & $T$~(ns) & $\error_{3}$
    \\
    \hline
    7049.654 & 6752.475 & 2180.4 & $0.8\%$
    \\
    7049.673 & 6761.585 & 3391.8 & $0.7\%$
    \\
    7049.699 & 6770.048 & 4158.8 & $1.9\%$
    \\
    7049.665 & 6793.903 & 4715.0 & $0.7\%$
    \\
    \hline
  \end{tabular*}
  \label{tab:3rdparams}
\end{table}

In order to analyze how well the effective theory of Eqs.~\eqref{eq:ham_eff_3}
and~\eqref{eq:gBS} describes the three-tone protocol shown in
Fig.~\ref{fig:pop}(b), we calculate $\op{g}_{\rmBS}$ for the parameters in
Table~\ref{tab:params} and the third drive in Eq.~\eqref{eq:3rdtone}. We find
$g_{\rmAB}/2\pi = - \SI{0.045}{\mega\hertz}$, $g_{\rma}^{\rmsb}/2\pi
= - \SI{0.54}{\mega\hertz}$, $g_{\rmb}^{\rmsb}/2\pi = - \SI{1.26}{\mega\hertz}$
and $\Delta/2\pi \approx - \SI{3}{\mega\hertz}$. This gives rise to an
effective beamsplitter interaction of $g_{\rmBS}/2\pi
= - \SI{0.27}{\mega\hertz}$~\footnote{We assume the transmon to be in the ground
state, thus replacing $\op{\sigma}_{\rmz}$ by $-1$ in Eq.~\eqref{eq:gBS}} and
thus to a SWAP time of $T \approx \SI{924}{\nano\second}$. We conjecture that
the discrepancy of the analytically predicted SWAP time with respect to the
numerically observed one of $T = \SI{1492}{\nano\second}$ is caused by
$|g_{\rma}^{\rmsb}|, |g_{\rmb}^{\rmsb}| \ll |\Delta|$ --- which needs to hold
for an accurate analytical predictions --- not being fulfilled sufficiently
well.

To verify this conjecture in more detail, we evaluate Eq.~\eqref{eq:gBS} --- and
the SWAP time it gives rise to --- for further three-tone protocols, where the
respective choice of the third frequency $\omega_{3}$ gives rise to larger
$|\Delta|$ such that $|g_{\rma}^{\rmsb}|, |g_{\rmb}^{\rmsb}| \ll |\Delta|$ is
better satisfied. This also allows to investigate the impact of $|\Delta|$ on
emerging cavity-transmon correlations and the protocol error, which is not
evident from the analytical treatment so far. Possible sets of parameters for
three-tone protocols can easily be found using gradient-free optimization as
described in Sec.~\ref{subsec:oct}. In these optimizations, we only allow
$\omega_{2}$ and $\omega_{3}$ to change in addition to $T$ and keep all three
amplitudes as well as $\omega_{1}$ fixed by their values in
Table~\ref{tab:params} and Eq.~\eqref{eq:3rdtone}. The reason behind this choice
is that $\omega_{3}$ primarily defines $\Delta$ while adapting $\omega_{2}$ is
required to correct for potential Stark shifts in Eq.~\eqref{eq:reson}. The
latter was assumed to hold while deriving Eqs.~\eqref{eq:ham_eff_3}
and~\eqref{eq:gBS}. Table~\ref{tab:3rdparams} presents a few optimization
results. We now use Eq.~\eqref{eq:gBS} to calculate $g_{\rmBS}$ and its
corresponding SWAP time for each set of parameters. In Fig.~\ref{fig:SWAP_times}
we compare the calculated SWAP times (black, solid line) with those obtained
numerically (green dots), cf.  Table~\ref{tab:3rdparams}, and with the
predictions of the semi-analytical method (red dots) developed in
Ref.~\cite{PRA.99.012314}. In the latter, the drives are treated
non-perturbatively using Floquet theory and the cavity-transmon couplings are
treated perturbatively using linear response theory.
Appendix~\ref{sec:app:floquet} summarizes the details of this method. We observe
a qualitative agreement between the methods and attribute the small remaining
discrepancies to the approximations made within each method.
Figure~\ref{fig:SWAP_times} suggests that Eqs.~\eqref{eq:ham_eff_3}
and~\eqref{eq:gBS} indeed provide the correct physical intuition for the
speed-up of the three-tone protocol compared to the original two-tone protocol.
In other words, the speed-up is due to exploiting sideband couplings between the
cavities and the transmon.

This explanation is also in agreement with the correlations emerging between
cavities and transmon under the three-tone protocol, cf. Fig.~\ref{fig:corr}(b),
as these correlations are not present under the original two-tone protocol, cf.
Fig.~\ref{fig:corr}(a). We observe a clear correspondence between the quantity
$\Delta$, cf. Eq.~\eqref{eq:gBS}, and the emergence of cavity-transmon
correlations. By comparing the correlation dynamics for all parameter sets of
Table~\ref{tab:3rdparams} (data not shown), we see a smooth transition from
behavior as in Fig.~\ref{fig:corr}(b) for the fastest three-tone protocol with
smallest $|\Delta|$, to behavior as in Fig.~\ref{fig:corr}(a) for the slowest
three-tone protocol with largest $|\Delta|$. This is also evident when
inspecting the coherent errors $\error_{3}$ of the three-tone protocols in
Table~\ref{tab:3rdparams}. Almost all sets of parameters give rise to coherent
protocol errors $\error_{3} < 1\%$ and are thus smaller than the coherent error
$\error_{3} = 2.6\%$ for three-tone protocol presented in
Fig.~\ref{fig:corr}(b). We see this as evidence that the cavity-transmon
correlations are mainly responsible for the coherent error $\error_{3}$ in the
three-tone protocol and ultimately prevent to find even faster protocol.

We conclude that, on one hand, small $|\Delta|$ is in general advantageous for
fast three-tone protocols, i.e., for $|g_{\rmBS}|$ to be large, cf.
Eq.~\eqref{eq:gBS}. On the other hand, $|\Delta|$ should not be chosen too small
such as to keep the coherent error due to non-vanishing correlations between
cavities and transmon at final time $T$ at bay. From a coherent perspective,
slower protocols are thus favorable. However, once decoherence is taken into
account, protocols should typically be as fast as possible. The optimal protocol
balances coherent error and the additional error from decoherence.

Finally, it should be mentioned that the calculations employing Floquet and
linear response theory --- used to determine the red dots in
Fig.~\ref{fig:SWAP_times} and detailed in Appendix~\ref{sec:app:floquet} ---
also provide an understanding of the emerging Kerr non-linearities caused by the
third drive. These non-linearities seem to increase significantly (data not
shown) compared to those induced by the original two-tone
protocol~\cite{PRX.8.021073}. Since the detrimental impact of such cavity
non-linearities for the SWAP operations becomes larger for higher photonic
states of the cavities, this might provide an explanation for the observed large
coherent protocol errors in case of qubit encodings involving higher photon
numbers. For a detailed study of drive-induced non-linearities of cavity modes
see Ref.~\cite{Zhang2021}.

\begin{figure}[tb!]
  \centering
  \includegraphics{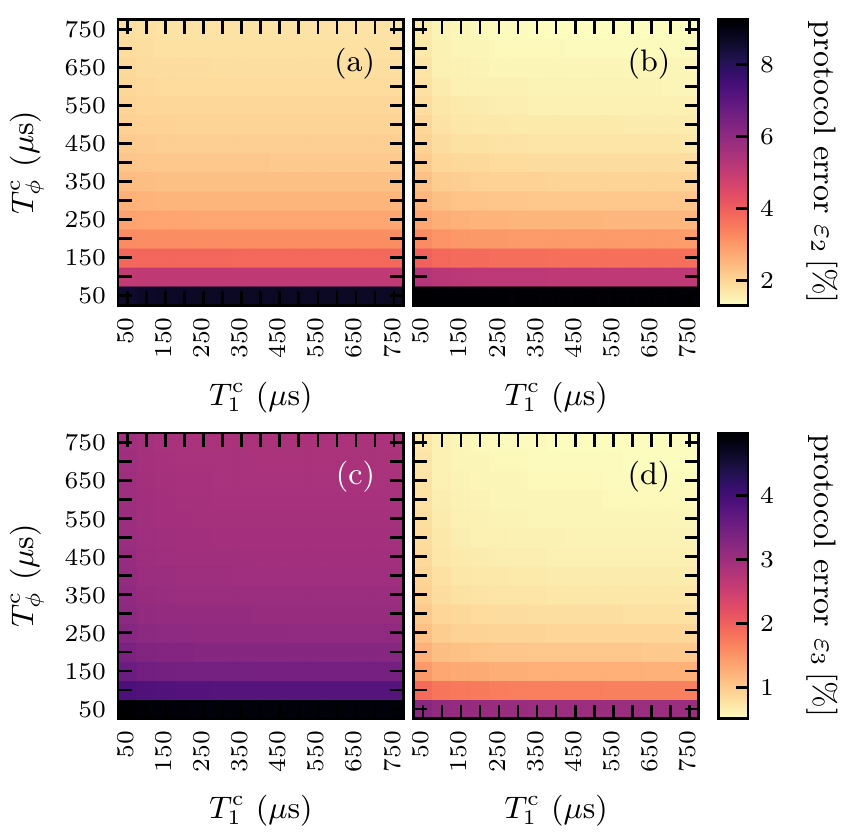}
  \caption{%
    Influence of decoherence on the error $\error_{2}$ of the original two-tone
    protocol (a) as well as its fine-tuned version (b) and the error
    $\error_{3}$ of the three-tone protocol (c) as well as its fine-tuned
    version (d). Note that the relaxation times, $T_{1}^{\rma}$ and
    $T_{1}^{\rmb}$, and pure dephasing times, $T_{\phi}^{\rma}$ and
    $T_{\phi}^{\rmb}$, for both cavities are fixed in all panels to those values
    given in Table~\ref{tab:params} and only the times for the transmon are
    varied.
  }
  \label{fig:liou_C_mod}
\end{figure}


\section{Prospects for High-Fidelity Protocols in the Presence of Noise}
\label{sec:diss}
We now study how the protocols will perform once the coherence times of
transmons become better. So far we have assumed the transmon relaxation and pure
dephasing times to be $T_{1}^{\rmc} = \SI{50}{\micro\second}$ and
$T_{\phi}^{\rmc} = \SI{50}{\micro\second}$. However, better devices exist
already with recently reported relaxation times up to $T_{1}^{\rmc}
= \SI{500}{\micro\second}$~\cite{Wang2022} and dephasing times up to
$T_{\phi}^{\rmc} = \SI{300}{\micro\second}$~\cite{Place2021}. In order to
investigate how the protocol errors change for better transmon devices, we
gradually increase both $T_{1}^{\rmc}$ and $T_{\phi}^{\rmc}$ from
$\SI{50}{\micro\second}$ to $\SI{750}{\micro\second}$. While the latter times
might still be out of reach for current devices, we consider them here to give
some perspective for possible future improvement. Figure~\ref{fig:liou_C_mod}(a)
and (b) show how the errors $\error_{2}$ and $\error_{2,\rmgrad}$ of the
original two-tone protocol and its fine-tuned version improve when
$T_{1}^{\rmc}$ and $T_{\phi}^{\rmc}$ increase. As can be seen, both errors show
a weak dependence on $T_{1}^{\rmc}$ (on the order of $\sim 0.1\%$) while they
rapidly decrease when increasing $T_{\phi}^{\rmc}$. For the recently reported
values of $T_{1}^{\rmc} = \SI{500}{\micro\second}$ and $T_{\phi}^{\rmc}
= \SI{300}{\micro\second}$, we find $\error_{2} = 2.5\%$ and $\error_{2,\rmgrad}
= 2.2\%$, which is further lowered to $\error_{2} = 1.8\%$ and
$\error_{2,\rmgrad} = 1.3\%$ for $T_{1}^{\rmc} = \SI{750}{\micro\second}$ and
$T_{\phi}^{\rmc} = \SI{750}{\micro\second}$.

In Fig.~\ref{fig:liou_C_mod}(c) and (d), we investigate how the errors
$\error_{3}$ and $\error_{3,\rmgrad}$ of the three-tone protocol and its
fine-tuned version scale. Similarly to the two-tone protocol, a weak
dependence on $T_{1}^{\rmc}$ is observed, while a larger $T_{\phi}^{\rmc}$
readily improves both $\error_{3}$ and $\error_{3,\rmgrad}$. For $T_{1}^{\rmc}
= \SI{500}{\micro\second}$ and $T_{\phi}^{\rmc} = \SI{300}{\micro\second}$, we
find $\error_{3} = 3.1\%$ and $\error_{3,\rmgrad} = 0.8\%$, which is further
lowered to $\error_{3} = 2.8\%$ and $\error_{3,\rmgrad} = 0.5\%$ for
$T_{1}^{\rmc} = \SI{750}{\micro\second}$ and $T_{\phi}^{\rmc}
= \SI{750}{\micro\second}$. These results are to be understood as follows. The
error of the constant three-tone protocol is essentially given by the remaining,
relatively large coherent error of $\error_{3} = 2.6\%$. It can not be lowered
below this value by solely improving coherence times. In contrast, for the
fine-tuned three-tone protocol an error of $<1\%$ is achievable even
for present day $T_{1}^{\rmc}$ and $T_{\phi}^{\rmc}$ times of the transmon. To
conclude, Fig.~\ref{fig:liou_C_mod} indicates that improving $T_{\phi}^{\rmc}$
yields the largest improvements in fidelity, in particular for the
fine-tuned protocols, while improving $T_{1}^{\rmc}$ has a rather small
effect. This may be explained by the transmon remaining in an energetically
low-lying Floquet state throughout the dynamics. This state is already very
close to the transmon's bare ground state and can thus not decay much further.
Moreover, it resembles a coherent state, which is naturally more resistant to
energy relaxation.

Our observation of the role of the Floquet state suggests a further possibility
to reduce the protocol's sensitivity to decoherence. It is motivated by
recognizing that the bare transmon ground state is not affected by any
relaxation or dephasing. Thus, it should be possible to engineer two- or
three-tone protocols that are less susceptible to transmon decoherence by
staying even closer to its bare ground state. In Appendix~\ref{sec:app:2tone} we
show that the ideal two-tone protocol that minimizes excitation of the transmon
--- which in fact minimizes the protocol error --- is achieved if $\xi_{1}
\approx \xi_{2}$ with $\xi_{k} = \Omega_{k}/\delta_{k}$ the normalized amplitude
and $\delta_{k} = \omega_{k} - \omega_{\rmc}$. Any deviation from $\xi_{1}
\approx \xi_{2}$ leads to more excitation of the transmon and thus larger
protocol errors. This insight on the normalized amplitude can serve as a guiding
principle for the design of further high fidelity two- and three-tone protocols.

\section{Conclusions and Outlook}
\label{sec:conclusions}
For the practical problem of engineering a beamsplitter interaction between two
bosonic modes by appropriately driving an intermediate coupling element, we have
shown how to use quantum optimal control theory (OCT) to systematically improve
performance and gain, at the same time, insight into the control mechanism. Key
was to combine a two-step optimization with comprehensive analysis of the
underlying dynamics, exploiting several available numerical and analytical
tools. In more detail, starting from an analytical two-tone
protocol~\cite{PRX.8.021073} that utilizes two drives with constant amplitudes
and frequencies, we have shown how, in a first step, a simple gradient-free
optimization technique can be used to enhance the beamsplitter strength by
roughly a factor of five. The increased strength originates from adding a third
tone with fixed amplitude and frequency. Our analysis revealed that the third
tone --- with its parameters determined by gradient-free optimization ---
induces and exploits near to resonant cavity-transmon sideband couplings to
strengthen the beamsplitter rate. The ability of our approach to identify this
rather non-intuitive solution to the considered control problem already
exemplifies the utility of OCT.

In a second step, we have then used a gradient-based optimization technique to
further improve the three-tone protocol identified by the gradient-free
optimization. This allows to further lower the protocol error. Remarkably, the
solutions identified this way are much simpler than solutions obtained with
gradient-based methods only. This is in accordance with Ref.~\cite{Goerz2015}
and emphasizes the advantage of a hybrid optimization approach --- combining
both gradient-free and gradient-based methods --- compared to any of the two
alone.

The improved beamsplitter strength of the three-tone protocol, obtained by the
gradient-free optimization, comes at the expense of introducing correlations
between cavities and transmon. These correlations do not vanish at times where
e.g.\ a SWAP gate should be implemented. The second step in the hybrid
optimization approach primarily acts to suppress the correlations at final time.
Other than that, it does not significantly change the three-tone protocol
identified in the first step of the hybrid optimization approach. The control
fields obtained this way are experimentally feasible at all stages of the hybrid
approach as the latter increases the complexity of the control problem only
stepwise and thus allows one to find overall simpler solutions.

The error for a SWAP gate significantly decreases due to the significant
increase in beamsplitter strength which leads to a reduction in protocol
duration and, as another consequence, diminished influence from decoherence. The
reduction in protocol time and error comes at the expense of making the
three-tone protocol more codeword-dependent than the original two-tone protocol,
i.e., dependent on the respective encoding of the qubits in the cavity Hilbert
space. Our findings nevertheless suggest that it is always possible to identify
optimized drives for a given encoding, as we have shown for the example of
binomial encoding~\cite{PRX.6.031006}. Whether a faster, codeword-agnostic
protocol exists, remains an open question.

Finally, the important interplay of OCT and the analytical tools, used to
identify and understand the obtained solutions, needs to be stressed. Together
with the analysis of the impact of decoherence onto the two- and three-tone
protocols it forms an ideal starting point for further improvements of the
protocols. For example, the insight that a symmetric choice of the normalized
amplitudes makes the protocol less susceptible to decoherence can be fed back
into the optimization to obtain even better protocols.

To summarize, we have shown how a specific protocol --- relevant in the context
of continuous-variable quantum computing --- can be accelerated and its error
minimized by means of OCT. This study therefore serves as a demonstration of how
OCT can be used systematically in order to solve or improve a given control
problem. In a first optimization step, a gradient-free optimization allows to
identify intelligible control strategies. They can be fine-tuned afterwards by
a gradient-based method in a second step to yield highly performant solutions,
while keeping the field shapes feasible. Since all OCT tools are readily
available~\cite{nlopt, SciPostPhys.7.6.080}, application of this procedure to
other problems of interest should be straightforward. Our analysis of the
optimization results and finding of a clear physical explanation for the
speed-up opens new avenues for further improvements of the beamsplitter
protocol.

\begin{acknowledgments}
  Financial support from the DAAD and the Deutsche Forschungsgemeinschaft (DFG),
  Project No. 277101999, CRC 183 (project C05), is gratefully acknowledged. The
  research of SMG and YZ was sponsored by the Army Research Office (ARO), and
  was accomplished under Grant No. W911NF-18-1-0212. The views and conclusions
  contained in this document are those of the authors and should not be
  interpreted as representing the official policies, either expressed or
  implied, of the Army Research Office (ARO), or the U.S. Government. The U.S.
  Government is authorized to reproduce and distribute reprints for Government
  purposes notwithstanding any copyright notation herein. DB would furthermore
  like to thank the Yale Quantum Institute for hospitality.
\end{acknowledgments}

\appendix

\section{Constructing an effective Hamiltonian for the two cavities}
\label{sec:app:construct}

\begin{figure*}[tb!]
  \centering
  \includegraphics{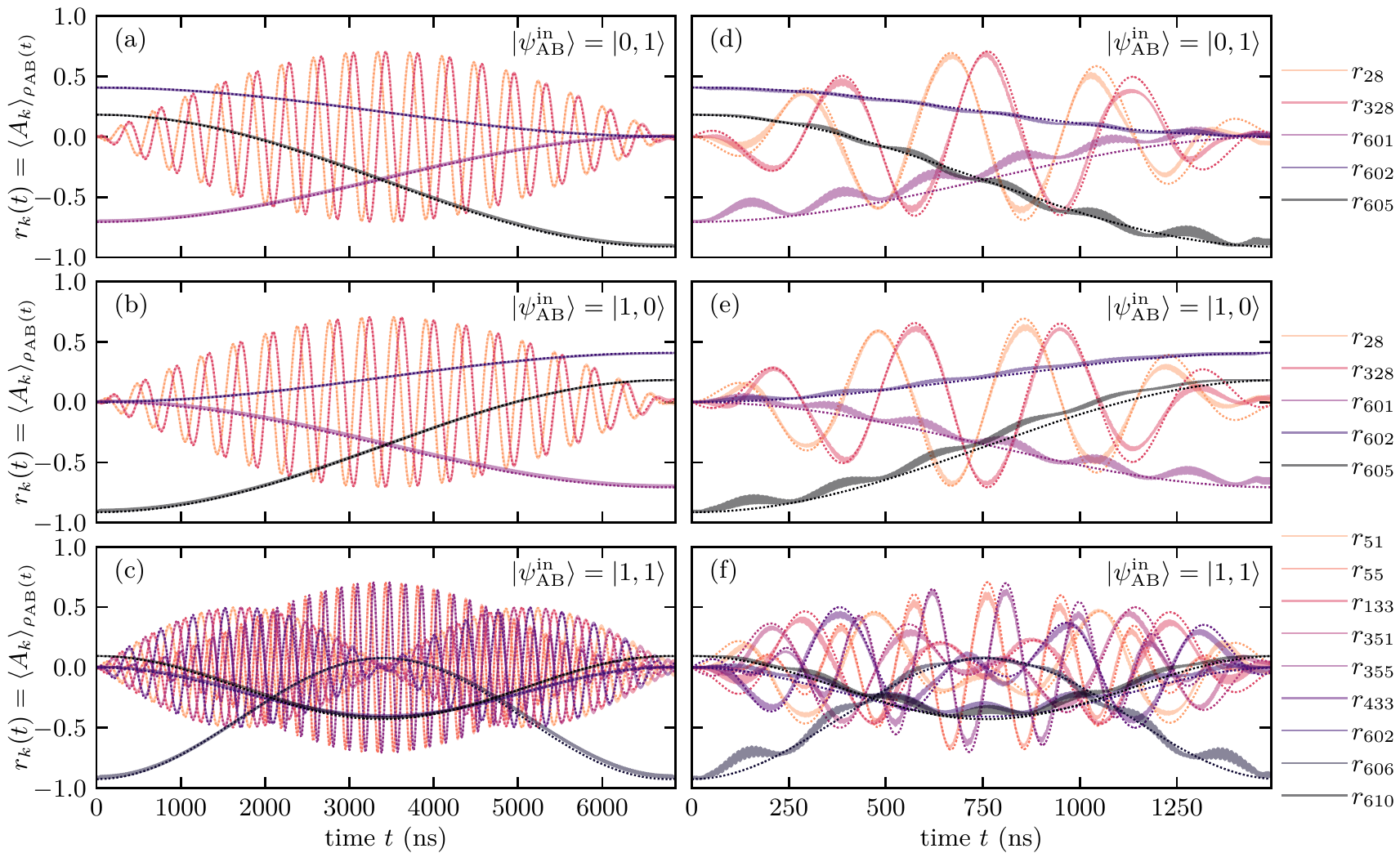}
  \caption{%
    Bloch trajectories for the reduced state of cavities A and B. Only those
    components $r_{k}(t) = \Braket{\op{A}_{k}}_{\op{\rho}_{\rmAB}(t)}$ in the
    generalized Bloch vectors are depicted that show significant changes in
    time. The opaque, solid lines correspond to the Bloch vector trajectories of
    $\op{\rho}_{\rmAB}(t)$ obtained by tracing out transmon C for the dynamics
    under the two-tone protocol (left column) and the three-tone protocol (right
    column) as shown in Fig.~\ref{fig:pop}. The dotted lines correspond to the
    Bloch vector trajectories obtained for a dynamics with the effective
    Hamiltonian~\eqref{eq:H_AB_eff}, which is only defined on the reduced system
    of the two cavities.
  }
  \label{fig:compare_bloch}
\end{figure*}

In this appendix we reconstruct --- based on the numerical data shown in
Fig.~\ref{fig:pop} --- an effective Hamiltonian that correctly describes the
dynamics of the reduced subsystem of the two cavities. As we will show, this
Hamiltonian works for both the two- and three-tone protocol confirming once more
that adding the third drive indeed gives rise to a stronger beamsplitter
interaction. This is particularly remarkable given the differences in the
correlation dynamics between Fig.~\ref{fig:corr}(a) and (b). The two-tone
protocol gives rise to almost unitary dynamics in the reduced subsystem of the
two cavities since the transmon stays uncorrelated at all times. In contrast,
the three-tone protocol gives rise to correlations between cavities and
transmon, cf. Fig.~\ref{fig:corr}, and thus indicates non-unitary dynamics of
the reduced subsystem of the cavities.

In the following, we inspect the reduced dynamics of the two cavities. To this
end, we introduce the generalized Bloch vector $\myvec{r}(t) = (r_{1}(t),
\dots, r_{M^{2}-1}(t))^{\top}$ of the reduced state $\op{\rho}_{\rmAB}(t)
= \tr_{\rmc}\{\op{\rho}(t)\}$ of cavities A and B,
\begin{align}
  \op{\rho}_{\rmAB}(t)
  =
  \frac{1}{M} \uop_{M} + \myvec{r}(t) \cdot \vecop{A},
\end{align}
where $M=\dim\{\hil_{\rma} \otimes \hil_{\rmb}\}$ and $\vecop{A} = (\op{A}_{1},
\dots, \op{A}_{M^{2}-1})^{\top}$. Here $\{\op{A}_{1}, \dots, \op{A}_{M^{2}-1}\}$
is a basis of traceless, Hermitian $M \times M$ matrices satisfying
$\braket{\op{A}_{i}, \op{A}_{j}} = \delta_{i,j}$. We choose the generalized
Gell-Mann matrices for this basis and order them according to their presentation
in Ref.~\cite{Bertlmann.JPhysAMathTheor.41.235303}. For the numerical
simulations presented in Fig.~\ref{fig:pop}, we have $M=25$, hence the
generalized Bloch vector $\myvec{r}(t)$ has $25^{2}-1=624$ components. For the
effective beamsplitter interaction of the two-tone protocol, cf.
Fig.~\ref{fig:pop}(a), many of these components are constant or almost constant
and only a small fraction shows a significant time-dependence. The opaque, solid
lines in Fig.~\ref{fig:compare_bloch}(a)-(c) show the dynamics of those
``relevant'' components for the initial states $\ket{\psi_{\rmAB}^{\mathrm{in}}}
\in \{\ket{0,1,0}, \ket{1,0,0}, \ket{1,1,0}\}$ in the bare basis.
Figure~\ref{fig:compare_bloch}(d)-(e) show the same components for the
three-tone protocol, cf. Fig.~\ref{fig:pop}(b). As can be seen, the slowly
changing components in Fig.~\ref{fig:compare_bloch}(d)-(e) follow a slightly
more complex version than their counterparts in panels (a)-(c) but are identical
in shape. In contrast, the rapidly oscillating components follow the same
envelope in both cases but their oscillations differ. Note the different time
scales of the dynamics. The rapid oscillations have actually the same frequency
for both the two- and three-tone protocol.

Since there is no immediate procedure to derive an effective Hamiltonian for the
subsystem of cavities A and B, we numerically find an effective Hamiltonian that
fits the dynamics observed in Fig.~\ref{fig:compare_bloch}. The effective
Hamiltonian reads
\begin{align} \label{eq:H_AB_eff}
  \op{H}_{\rmeff}^{\rmAB}(t)
  =
  -
  g_{\rmBS}
  \left(%
    e^{- \im \omega_{\rmz} t}
    \op{a} \op{b}^{\dagger}
    +
    e^{  \im \omega_{\rmz} t}
    \op{a}^{\dagger} \op{b}
  \right)
  -
  \frac{\omega_{\rmz}}{2}
  \left(%
    \op{a}^{\dagger} \op{a}
    -
    \op{b}^{\dagger} \op{b}
  \right).
\end{align}
For the original two-tone protocol, the two parameters of the effective
Hamiltonian~\eqref{eq:H_AB_eff} are $g_{\rmBS}/2\pi = \SI{36.9}{\kilo\hertz}$
and $\omega_{\rmz}/2\pi = \SI{2.625}{\mega\hertz}$. For the three-tone
protocol, the two parameters are $g_{\rmBS}/2\pi = \SI{165}{\kilo\hertz}$ and
$\omega_{\rmz}/2\pi = \SI{2.625}{\mega\hertz}$. In both cases, the parameters
are obtained by fitting the effective, analytical curves generated by
Hamiltonian~\eqref{eq:H_AB_eff} to the numerical curves of
Fig.~\ref{fig:compare_bloch}. As expected, the effective beamsplitter
interaction $g_{\rmBS}$ is larger for the three-tone protocol. It increases by
a factor $4.47$ which roughly matches the decrease in protocol duration (factor
$4.54$). Interestingly, the rapidly oscillating components in
Fig.~\ref{fig:compare_bloch} can be reproduced by the same frequency
$\omega_{\rmz}/2\pi = \SI{2.625}{\mega\hertz}$ for both protocols. It
exactly matches the relative Stark shift for $\omega_{a}$ and $\omega_{b}$ in
the original two-tone protocol, which can be calculated from the parameters in
Table~\ref{tab:params} and Eq.~\eqref{eq:reson} via
\begin{align}
  (\omega_{2} - \omega_{1}) - (\omega_{\rmb} - \omega_{\rma})
  =
  \SI{2.625}{\mega\hertz}.
\end{align}
This readily explains the difference between Eq.~\eqref{eq:HabBS} and
Eq.~\eqref{eq:H_AB_eff}, as the latter describes the dynamics in the frame set
by Hamiltonian~\eqref{eq:ham'}, i.e., in a rotating frame where both cavities
A and B have vanishing level splittings $\omega_{\rma}$ and $\omega_{\rmb}$.
However, this frame does not capture the Stark shifts $\omega_{\rma} \rightarrow
\som_{\rma}$ and $\omega_{\rmb} \rightarrow \som_{\rmb}$ induced by the drives
on the transmon. In consequence, both cavities have still non-vanishing level
splittings in the rotating frame and hence the non-vanishing $\omega_{\rmz}$ in
Eq.~\eqref{eq:H_AB_eff}.

The fact that $\omega_{\rmz}$ is identical for the two- and three-tone protocol
is surprising, since the third drive could, in principle, give rise to different
individual Stark shifts of $\omega_{\rma}$ and $\omega_{\rmb}$. However, the
relative Stark shift is identical for both protocols. This might be viewed as
another numerical confirmation of the effective theory presented in
Eqs.~\eqref{eq:ham_eff_3} and~\eqref{eq:gBS}, namely that the first two tones
are exclusively responsible for activating the beamsplitter interaction via
Eq.~\eqref{eq:reson} while the third tone exclusively adds the cavity-transmon
sideband transitions.

\section{Floquet calculation of the beamsplitter rate}
\label{sec:app:floquet}
The Floquet results shown in Fig.~\ref{fig:SWAP_times} of the main text were
obtained by using the method developed in Ref.~\cite{PRA.99.012314}, which we
briefly describe here. Note that the following Floquet treatment is an
approximate method to estimate the beamsplitter rate.

As a first step, we start from the Hamiltonian in Eq.~\eqref{eq:ham} and switch
to the rotating frame at the frequency $\omega_{1}$ of driving field 1. This
leads to the Hamiltonian
\begin{align} \label{eq:H}
  \op{H}(t)
  &=
  \op{H}_{\rmc}(t)
  +
  (\omega_{\rma} - \omega_{1}) \op{a}^{\dagger} \op{a}
  +
  (\omega_{\rmb} - \omega_{1}) \op{b}^{\dagger} \op{b}
  \notag \\
  &\quad
  +
  g_{\rma} \left(\op{a} \op{c}^{\dagger} + \op{a}^{\dagger} \op{c}\right)
  +
  g_{\rmb} \left(\op{b} \op{c}^{\dagger} + \op{b}^{\dagger} \op{c}\right),
  \\
  \op{H}_{\rmc}(t)
  &=
  (\omega_{\rmc} - \omega_{1}) \op{c}^{\dagger} \op{c}
  -
  \frac{\alpha_{\rmc}}{2} \op{c}^{\dagger} \op{c}^{\dagger} \op{c} \op{c}
  \notag \\
  &\quad
  +
  \left(%
    \Omega_{1}
    +
    \Omega_{2} e^{-\im \omega_{21} t}
    +
    \Omega_{3} e^{-\im \omega_{31} t}
  \right) \op{c}^{\dagger}
  \notag \\
  &\quad
  +
  \left(%
    \Omega_{1}^{*}
    +
    \Omega_{2}^{*} e^{\im \omega_{21} t}
    +
    \Omega_{3}^{*} e^{\im \omega_{31} t}
  \right) \op{c}.
  \label{eq:H_c}
\end{align}
Here we consider time-independent drive amplitudes $\Omega_{1,2,3}$, and we
define $\omega_{21}$ and $\omega_{31}$ to be $\omega_{21} = \omega_2
- \omega_1$, $\omega_{31} = \omega_3 - \omega_1$. 

The key idea of the method is to treat the cavity-transmon couplings in
Eq.~\eqref{eq:H} above perturbatively, but treat the drives non-perturbatively
using Floquet theory. To apply Floquet theory, we require $\omega_{21}$ to be
commensurate with $\omega_{31}$. In practice, this is done by keeping finite
number of digits for the numerical values of the drive frequencies. In the
results shown in Fig.~\ref{fig:SWAP_times}, we round the drive frequencies in
unit of MHz to the closest integer. For instance, $\omega_{2}/2\pi$ is rounded
to $\SI{7050}{\mega\hertz}$. After this, we find the smallest positive integers
$p$ and $q$ such that the ratio of $\omega_{21}$ and $\omega_{31}$ is given by
$p/q$. Then the transmon Hamiltonian $\op{H}_{\rmc}$ in Eq.~\eqref{eq:H_c} is
time-periodic, namely,
\begin{align}
  \op{H}_{\rmc}(t+2\pi/\omega_{\mathrm{H}})
  =
  \op{H}_{\rmc}(t),
  \quad
  \omega_{\mathrm{H}}
  =
  \omega_{21}/p = \omega_{31}/q.
\end{align}
Because $\op{H}_{\rmc}(t)$ is periodic in time, by Floquet theorem, there is
a set of Floquet eigenstates associated with $\op{H}_{\rmc}(t)$ --- analog to
stationary eigenstates for static Hamiltonians. These Floquet states can be
written in the form
\begin{align}
  \psi_{m}(t)
  &=
  e^{-\im \epsilon_{m} t} u_{m}(t),
\end{align}
where $\epsilon_{m}$ is the quasienergy and $u_{m}$ is called the Floquet mode
which has the same periodicity as the Hamiltonian, i.e.,
$u_{m}(t+2\pi/\omega_{\mathrm{H}}) = u_{m}(t)$. 

As derived in Ref.~\cite{PRA.99.012314}, to leading order in the coupling
strengths $g_{\rma}, g_{\rmb}$, the cavity-cavity beamsplitter rate when the
transmon is in the $m$th Floquet state is given by the following formula
\begin{align} \label{eq:g_BS}
  g_{\rmBS,m}
  &=
  g_{\rma}^{*} g_{\rmb} \sum_{n} \sum_{K}  \left[%
    \frac{%
      \op{c}_{mn,K+1} (\op{c}^{\dagger})_{nm,-K}
    }{%
      \omega_{\rmb} - \omega_{1} + K \omega_{\mathrm{H}} + \epsilon_{mn}
    }
    \right.
    \notag \\
    &\quad +
    \left.
    \frac{%
      (\op{c}^{\dagger})_{mn,-K} \op{c}_{nm,K+1}
    }{%
      - \omega_{\rmb} + \omega_{1} - K \omega_{\mathrm{H}} + \epsilon_{mn}
    }
  \right], 
\end{align}
where $\epsilon_{mn} = \epsilon_{m} - \epsilon_{n}$ and where $\op{c}_{mn,K}$ is
the $K$th Fourier component of the matrix element of the transmon operator
$\op{c}$ between its Floquet modes $u_{m}$ and $u_{n}$,
\begin{align}
  \op{c}_{mn,K}
  =
  \frac{\omega_{\mathrm{H}}}{2\pi}
  \int_{0}^{2\pi/\omega_{\mathrm{H}}}
  \langle u_{m}(t)| \op{c} | u_{n}(t)\rangle
  e^{-\im K \omega_{\mathrm{H}} t} \dd t.
\end{align}

To obtain the Floquet results in Fig.~\ref{fig:SWAP_times}, we set $m=0$ in
Eq.~\eqref{eq:g_BS}, which corresponds to the Floquet state that adiabatically
connects to the transmon ground state without the drive. Near the sideband
resonance, we find that using the dressed frequency of cavity B ---
approximately given by $\omega_{\rmb} + g_{\rmb}^2
/ (\omega_{\rmb}-\omega_{\rmc})$ --- in place of bare frequency
$\omega_{\rmb}$ in Eq.~\eqref{eq:g_BS} produces a better agreement with the
time-domain numerical results. To obtain the error bars in
Fig.~\ref{fig:SWAP_times}, we simply shift $\omega_{\rmb}/2\pi$ by $\pm
\SI{1}{\mega\hertz}$. This allows us to explore the sensitivity of the
beamsplitter rate on the distance to the sideband resonance. 

\section{Constructing two- and three-tone protocols with minimal transmon
excitation}
\label{sec:app:2tone}

\begin{figure}[tb!]
  \centering
  \includegraphics{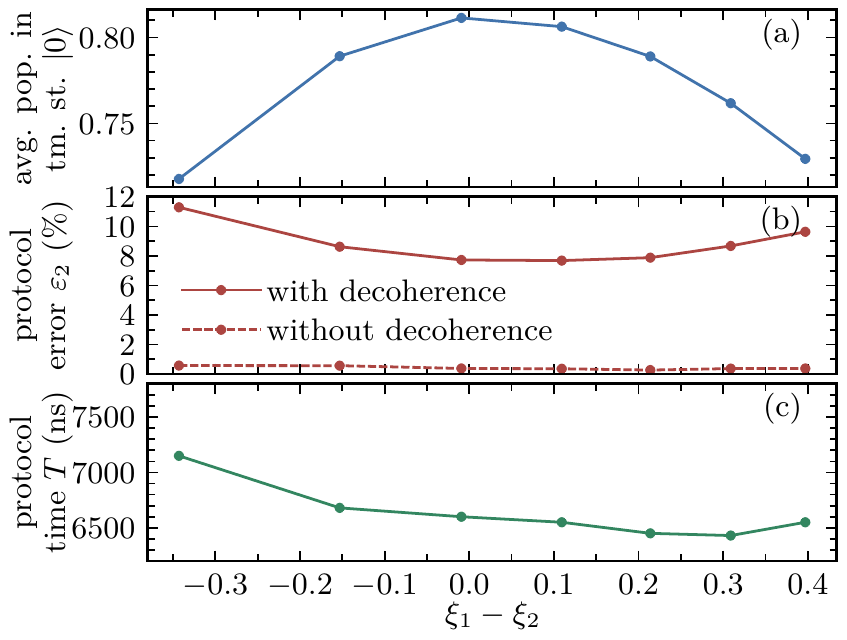}
  \caption{%
    Error analysis for various two-tone protocols based on the difference
    $\xi_{1} - \xi_{2}$ of the normalized amplitudes $\xi_{1}$ and $\xi_{2}$ but
    conditioned by $\xi_{1} \xi_{2} = 0.126$.
    Panel (a) shows the average bare ground state population of the transmon
    (without decoherence) for the respective protocol, cf.
    Eq.~\eqref{eq:avg_pop_tm_0}.
    Panel (b) shows the protocol error $\error_{2}$ with and without
    decoherence, while (c) shows the corresponding protocol duration $T$.
  }
  \label{fig:2tone_ana}
\end{figure}

In this appendix, we examine how to engineer two- and three-tone protocols with
minimal excitation of the transmon. While one might suppose that this only helps
in improving robustness with respect to $T_{1}^{\rmc}$, recall that with fewer
excitations in higher transmon levels, also fewer coherences with respect to
these levels occur. To identify such protocols, we need to find drive parameters
for which the Floquet state --- to which the transmon is dynamically transferred
by switching the drive on and off --- is closest to the bare ground state. To
this end, we compare various two-tone protocols under the constraint of
identical $|\xi_{1} \xi_{2}|$, where $\xi_{k} = \Omega_{k}/\delta_{k}$ is the
normalized amplitude and $\delta_{k} = \omega_{k} - \omega_{\rmc}$ a detuning.
Note that despite $|\xi_{1} \xi_{2}|$ being the main quantity that defines the
beamsplitter rate $g_{\rmBS}$, cf. Eq.~\eqref{eq:HabBS}, and thus the protocol
duration $T$, the individual physical amplitudes $\Omega_{1}$ and $\Omega_{2}$
and thus individual normalized amplitudes $\xi_{1}$ and $\xi_{2}$ can still be
chosen differently.

Figure~\ref{fig:2tone_ana}(a) shows the average bare ground state population of
the transmon, defined via
\begin{align} \label{eq:avg_pop_tm_0}
  \frac{1}{T M^{2}} \sum_{\na,\nb=0}^{M-1} \int_{0}^{T} \dd t
  \lol{%
    \ket{0}\bra{0} \bigg|
    \tr_{\mathrm{ab}}\left\{
      \dmap_{t,0} \left[\op{\rho}_{\na,\nb,0}\right]
    \right\}
  },
\end{align}
as a function of the difference $\xi_{1} - \xi_{2}$ for the two-tone protocol.
As can be seen, the average transmon ground state population is maximal if
$\xi_{1} - \xi_{2} \approx 0$, i.e., both drives have roughly the same
normalized amplitude. All these two-tone protocols have coherent errors $<
0.6\%$, see dashed line in Fig.~\ref{fig:2tone_ana}(b). Once decoherence is
taken into account, the protocols with larger average transmon ground state
population have smaller protocol errors $\error_{2}$ (solid line). The minimal
error roughly occurs for $\xi_{1} - \xi_{2} = 0$. Due to the constraint of
identical $|\xi_{1} \xi_{2}|$, the protocol durations $T$ are very similar
(albeit not identical), cf. Fig.~\ref{fig:2tone_ana}(c).

While Fig.~\ref{fig:2tone_ana} shows results for one particular choice of
$|\xi_{1} \xi_{2}|$, one might conjecture that weaker values --- and thus
longer protocol durations --- lead to larger average transmon ground state
population and hence more robustness with respect to transmon decoherence.
However, our observations indicate that being faster is always advantageous.
Thus, in order to find the protocol with best resistance against decoherence,
the primary goal should be to be fast and the secondary goal should be to stay
on average as close as possible to the bare transmon ground state. This
statement should hold for any three-tone protocol as well, since the average
bare ground state population decreases only slightly once the third drive is
turned on, decreasing for instance from $0.73$ to $0.70$ for the protocols
discussed in Fig.~\ref{fig:pop}. Figure~\ref{fig:2tone_ana} thus represents
a good starting point to find suitable two-tone protocols that can subsequently
be turned into high fidelity three-tone protocols.


%

\end{document}